\documentclass[12pt,preprint2]{aastex}
\usepackage{amsmath, mathrsfs}
\begin{document}

\newcommand{\sgn}{\operatorname{sgn}}
\newcommand{\kepler}{{\it Kepler}}
\newcommand{\rearth}{\ensuremath{R_{\oplus}}}
\newcommand{\rsun}{\ensuremath{R_\sun}}

\title{Detection of Potential Transit Signals in 17 Quarters of \textit{Kepler} Mission Data}

\author{Shawn Seader\altaffilmark{1}, Jon M. Jenkins\altaffilmark{2}, Peter Tenenbaum\altaffilmark{1}, Joseph D. Twicken\altaffilmark{1}, Jeffrey C. Smith\altaffilmark{1}, Rob Morris\altaffilmark{1}, Joseph Catanzarite\altaffilmark{1}, Bruce D. Clarke\altaffilmark{1}, Jie Li\altaffilmark{1}, Miles T. Cote\altaffilmark{2}, Christopher J. Burke\altaffilmark{1},  Sean McCauliff\altaffilmark{3}, Forrest R. Girouard\altaffilmark{3}, Jennifer R. Campbell\altaffilmark{3}, Akm Kamal Uddin\altaffilmark{3}, Khadeejah A. Zamudio\altaffilmark{3}, Anima Sabale\altaffilmark{3}, Christopher E. Henze\altaffilmark{2}, Susan E. Thompson\altaffilmark{1}, Todd C. Klaus\altaffilmark{4} } 
\altaffiltext{1}{SETI Institute/NASA Ames Research Center, Moffett Field, CA 94035, USA}
\altaffiltext{2}{NASA Ames Research Center, Moffett Field, CA 94035, USA}
\altaffiltext{3}{Wyle Labs/NASA Ames Research Center, Moffett Field, CA 94035, USA}
\altaffiltext{4}{Moon Express, Inc., P.O.Box 309, Moffett Field, CA 94035, USA}
\email{shawn.seader@nasa.gov}


\begin{abstract}
We present the results of a search for potential transit signals in the full $17$-quarter data set collected during \textit{Kepler}'s primary mission that ended on May 11, 2013, due to the on-board failure of a second reaction wheel needed to maintain high precision, fixed, pointing.  The search includes a total of $198,646$ targets, of which $112,001$ were observed in every quarter and $86,645$ were observed in a subset of the $17$ quarters.  \textbf{For the first time, this multi-quarter search is performed on data that have been fully and uniformly reprocessed through the newly released version of the Data Processing Pipeline.} We find a total of $12,669$ targets that contain at least one signal that meets our detection criteria: periodicity of the signal, a minimum of three transit events, an acceptable signal-to-noise ratio, and four consistency tests that suppress \textbf{many} false positives.  Each target containing at least one transit-like pulse sequence is searched repeatedly for other signals that meet the detection criteria, indicating a multiple planet system.  This multiple planet search adds an additional $7,698$ transit-like signatures for a total of $20,367$.  Comparison of this set of detected signals with a set of known and vetted transiting planet signatures in the \textit{Kepler} field of view shows that the recovery rate of the search is $90.3$\%.  We review ensemble properties of the detected signals and present various metrics useful in validating these potential planetary signals.  We highlight previously undetected \textbf{transit-like signatures, including several that may represent small objects} in the habitable zone of their host stars.
\end{abstract}

\section{Introduction}
\label{s:intro}
We have reported on the results of past searches of the \textit{Kepler} data for transiting planet signals in \citet{pt2012} which searched $3$ quarters of data, \citet{pt2013} which searched $12$ quarters of data, and \citet{pt2014} which searched $16$ quarters of data.  We now update and extend those results to incorporate the full data set collected by \textit{Kepler} and an additional year of \textit{Kepler} Science Processing Pipeline \citep{jenkins2010} development.  We further extend the results to include some metrics used to validate the astrophysical nature of the detections.  

\subsection{\textit{Kepler} Science Data}
\label{ss:data}
The details of \kepler{} operation and data acquisition have been reported elsewhere \citep{science-ops}.
In brief:  the \kepler{} spacecraft is in an Earth-trailing heliocentric orbit and maintained a boresight pointing centered 
on $\alpha = 19^{\rm h}22^{\rm m}40^{\rm s}, \delta = +44.5\degr$ during the primary mission.  The \kepler{} photometer acquired data on a 115 square degree region of the sky.  The data
were acquired in 29.4 minute integrations, colloquially known as ``long cadence'' data.  The spacecraft was rotated about its boresight axis by 90 degrees every 93 days in order to keep its solar panels and thermal
radiator correctly oriented; the interval which corresponds to a particular rotation state is known colloquially 
as a ``quarter.'' Because of the quarterly rotation, target stars were observed throughout the year in 4 different
locations on the focal plane. Science acquisition was interrupted monthly for data downlink, quarterly for
maneuver to a new roll orientation (typically this is combined with a monthly downlink to limit the loss of 
observation time), once every 3 days for reaction wheel desaturation (one long cadence sample is sacrificed
at each desaturation), and at irregular intervals due to spacecraft anomalies. In addition to these interruptions
which were required for normal operation, data acquisition was suspended for 11.3 days, from 2013-01-17 19:39Z
through 2013-01-29 03:50Z (555 long cadence samples)
\footnote{Time and date are presented here in ISO-8601 format, YYYY-MM-DD HH:MM, or optionally 
YYYY-MM-DD HH:MM:SS, with a trailing `Z' to denote UTC.} during this time, the spacecraft reaction wheels
were commanded to halt motion in an effort to mitigate damage which was being observed on reaction wheel
4, and spacecraft operation without use of reaction wheels is not compatible with high-precision photometric
data acquisition.

In July 2012, one of the four reaction wheels used to maintain spacecraft pointing during science acquisition
experienced a catastrophic failure. The mission was able to continue using the remaining three wheels to 
permit 3-axis control of the spacecraft, until May of 2013. At that time a second reaction wheel failed, forcing an end to \kepler{} data acquisition in the nominal \kepler{} field of view. As a result, the analysis reported
here is the first which incorporates the full volume of data acquired from that field of view.\footnote{We exclude 10 days of data acquired at the end of commissioning on $\sim$53,000 stars dubbed Q0 as this segment of data is too short to avoid undesirable edge effects in the transit search.}

\kepler{} science data acquisition of Quarter 1 began at 2009-05-13 00:01:07Z, and acquisition of Quarter 17 data
concluded at 2013-05-11 12:16:22Z. This time interval contains 71,427 long cadence intervals.  Of these,
5,077 were consumed by the interruptions listed above.  An additional 1,100 long cadence intervals were excluded from use 
in searches for transiting planets. These
samples were excluded due to data anomalies which came to light during processing and inspection of the 
flight data.  This includes a contiguous set of 255 long cadence samples acquired over the 5.2 days which immediately 
preceded the 11 day downtime described above: the shortness of this dataset combined with the duration of the
subsequent gap led to a judgement that the data would not be useful for transiting planet searches. A total of 65,250
 long cadence intervals for each target were dedicated to science data acquisition.  This is only 1,242 or $\sim$2\%  more searchable cadences 
than available for the analysis of 16 quarters of kepler data in \citet{pt2014}. This highlights the fact that quarter 17 was only about a
month long in duration before the failure of the second reaction wheel. 

\begin{figure}
\plotone{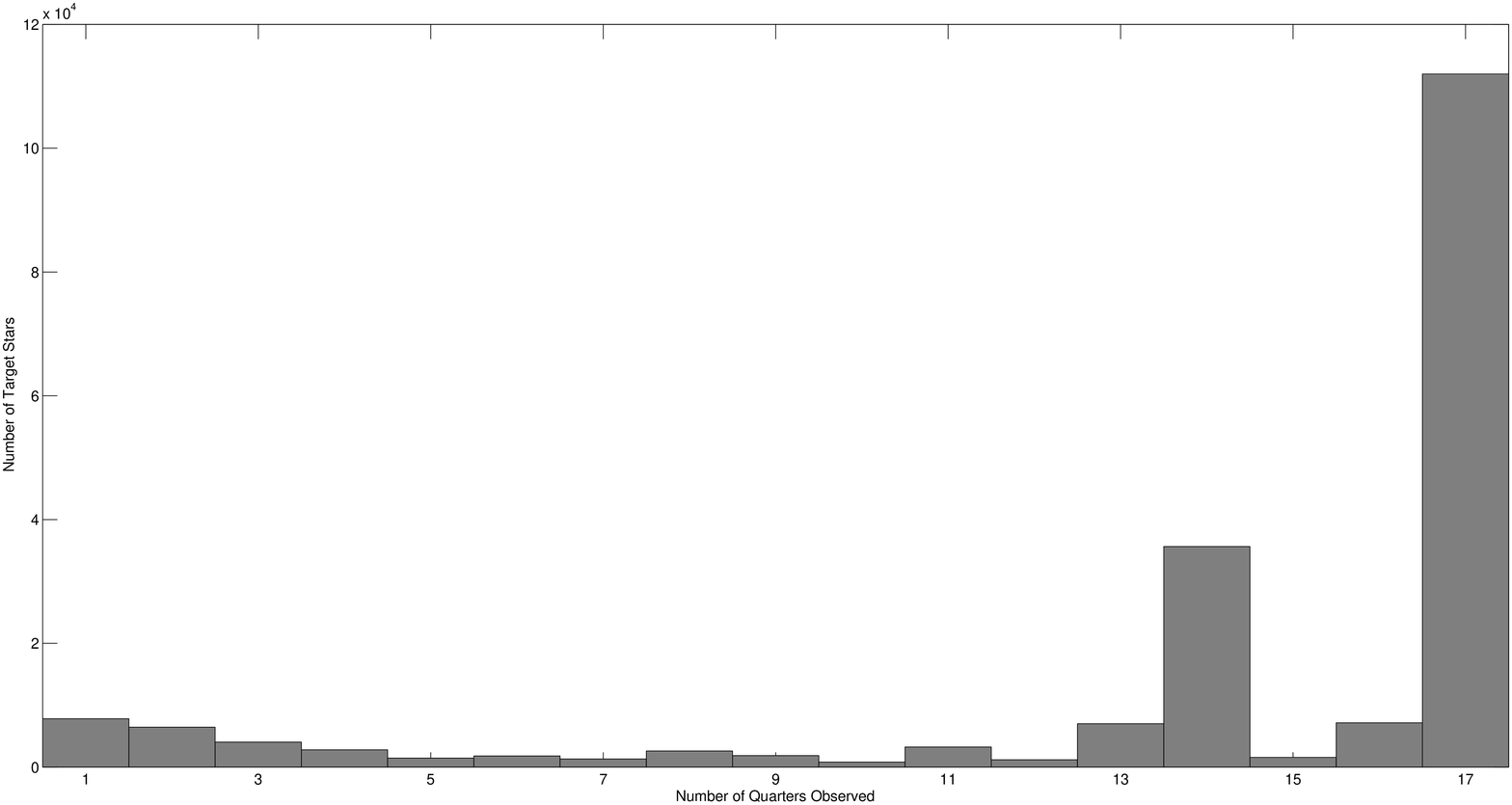}
\caption{Histogram of the number of quarters observed for all targets.
\label{f1}}
\end{figure}

A total of 198,675 targets observed by \kepler{} were searched for evidence of transiting planets. This set of targets
includes all stellar targets observed by \kepler{} at any point during the mission, and specifically includes target
stars which were not originally observed for purposes of transiting planet searches (asteroseismology targets, guest
observer targets, etc.). The exception to this is a subset of known eclipsing binaries, as described below.  
Figure \ref{f1} shows the distribution of targets according
to the number of quarters of observation. A total of 112,014 targets were observed for all 17 quarters. An additional
35,653 targets were observed for 14 quarters: the vast majority of these targets were in regions of the sky which
are observed in some quarters by CCD Module 3, which experienced a hardware failure in its readout electronics
during Quarter 4 in January 2010, resulting in a ``blind spot'' which rotates along with the \kepler{} spacecraft, gapping 25\% of the quarterly observations for affected targets.  The balance of 51,008
targets observed for some other number of quarters is largely due to gradual changes in the target
selection process over the duration of the mission.

As described in \citet{pt2014}, some known eclipsing binaries were excluded from planet searches in the pipeline. For this search over Q1-Q17, a total of
1,033 known eclipsing binaries were excluded. This is smaller than the number excluded in the Q1-Q16
analysis due to a change in exclusion criteria. Specifically, we excluded eclipsing binaries from the most recent 
Kepler catalog of eclipsing binaries \citep{eb-cat1,eb-cat2} that did not have a morphology $<$ 0.6 \citep{morph}, 
i.e., the primary and secondary eclipses must be well separated from one another, and were not in conflict with established planet candidates archived at NExScI.  This requirement on the morphology is identical
to that of the Q1-Q16 run but all the other criteria have been dropped to exclude fewer targets.  Note, however, that the Data
Validation (DV) \citep{hw2010,jt2014} step of the Data Processing Pipeline may still exclude some events based on its own analysis of 
primary and secondary depths.  Thus, the excluded eclipsing binaries are largely
contact binaries which produce the most severe misbehavior in the Transiting Planet Search (TPS)
pipeline module \citep{pt2012,pt2013}, 
while well-detached, transit-like, eclipsing binaries
are now processed in TPS.  This was done in order to ensure that no possible transit-like signature was excluded, 
and also to produce examples of the outcome of processing
such targets through both TPS and DV, such that quantitative differences
between planet and eclipsing binary detections could be determined and exploited for rejecting other,
as-yet-unknown, eclipsing binaries detected by TPS.

\subsection{Processing Sequence: Pixels to TCEs to KOIs}

The steps in processing \kepler{} science data have not changed since \citet{pt2012}, and are briefly summarized below.
The pixel data from the spacecraft were first calibrated, in the CAL pipeline module, to remove pixel-level effects such as gain variations, linearity, 
and bias\citep{calSpie2010}. The calibrated pixel values were then combined, within each target, in the Photometric Analysis (PA) pipeline module, to produce a flux time series for that target \citep{jdt2010b}.
In the Pre-search Data Conditioning (PDC) pipeline module the ensemble of target flux time series were then corrected for systematic variations driven by effects such as 
differential velocity aberration, temperature-driven focus changes, and small instrument pointing excursions\citep{stumpeMSMAP2014}. These corrected flux
time series became the inputs for the Transiting Planet Search.

The Transiting Planet Search software module analyzed each corrected flux time series individually for evidence of
periodic reductions in flux which would indicate a possible transiting planet signature. The search process incorporated
a significance threshold against a multiple event statistic"\footnote{The multiple event statistic is a measure of the degree to which the data are correlated with the reference waveform (in this case a sequence of evenly spaced transit pulses) normalized by the strength of the observation noise. It is approximately the same as the result of dividing the fitted transit depth by the uncertainty in the fitted transit depth, and can be interpreted in terms of the likelihood a value would be seen at that level by chance.}
and a series of vetoes; the latter were necessary because while the significance threshold was sufficient
for rejection of the null hypothesis, it was incapable of discriminating between multiple competing alternate hypotheses
which can potentially explain the flux excursions. An ephemeris on a given target which satisfied the significance threshold
and passed all vetoes is known as a Threshold Crossing Event (TCE). Each target with a TCE was then searched for
additional TCEs, which potentially indicated multiple planets orbiting a single target star.

After the search for TCEs concluded, additional automated tests were performed to assist members of the Science Team 
in their efforts to reject false positives. A TCE which has been accepted as a valid astrophysical signal (either planetary or an eclipsing binary), based on analysis of these
additional tests, is designated as a Kepler Object of Interest (KOI).  The collection of KOIs that pass additional scrutiny are promoted to planet candidate status, while those that don't are dispositioned as false positives.  Note that, while the TCEs were determined in a purely algorithmic
fashion by the TPS software module, KOIs were selected on the basis of examination and analysis by scientists.

\subsection{Pre-Search Processing}

Since the publication of \citet{pt2014}, there have been considerable improvements to the Kepler Science Processing Pipeline. 
We describe here only the recent improvements and refer the interested reader to \citet{pt2012,pt2013,pt2014} 
for a description of past improvements.  The pixel-level calibration performed by the CAL module of the 
Science Processing Pipeline has been improved to include a cadence-by-cadence 2-D black correction.  In an effort to mitigate the 
effects of the image artifacts described in \citet{caldwell} on the transiting planet search, 
CAL also generates rolling band flags \citep{RBA} that can be used to identify the cadences on a given target that are affected.
Improvements to the undershoot correction have also been made in CAL.  

The Pre-search Data Conditioning (PDC) module corrects for both attitude tweaks and Sudden Pixel Sensitivity Dropouts (SPSDs).
Previously it was noticed that those corrections can occasionally be very poor for several reasons such as local trends or noise 
in the data.  Improvements were made to fix this problem of poor corrections.  Additionally, some improvements were made to the wavelet band-splitting method used by PDC's multi-scale Bayesian Maximum A Posteriori algorithm for removing systematic noise from the data 
\citep{js2012}.

In addition to improvements in the science algorithms used to process the data, the data was also improved.  A major 
effort was undertaken by the Kepler Science Operations staff to, for the first time, uniformly reprocess all the data 
through the newly released codebase.  This was aided by the fact that all major modules of the pipeline have been ported 
over to the NASA Ames supercomputer to allow for reprocessing of multiple quarters in parallel. In all however, processing all the data uniformly from the start to the end of the 
pipeline takes several months.  Each module of the pipeline is necessarily, and painstakingly, configured and managed separately 
as they are run in sequence.

Pixel-level data as well as light curves from these pre-search processing steps are publicly available at the Mikulski Archive for Space Telescopes\footnote{
https://archive.stsci.edu/index.html} (MAST)
as Data Release 24.  Each data release is also accompanied by a Data Release Notes document that describes the data in general.  

\section{Transiting Planet Search}

This section describes the changes which have been made to the TPS algorithm since \citet{pt2014}.  
For further information on the algorithm, see \citet{jmj2002}, \citet{jmj2010}, \citet{pt2012,pt2013,pt2014}.

\subsection{Conditioning and Quarter-Stitching the Data}\label{detrend}
Owing to the complicated and varied noise sources across the many stellar targets in the \kepler{} field, 
the core detection algorithm in TPS is a wavelet-based, adaptive matched filter.  The data are transformed to the wavelet 
domain using Daubechies 12-tap wavelets \citep{debauchies} in a joint time-frequency overcomplete wavelet decomposition. 
In the wavelet domain, the noise at each time-frequency scale is estimated and used to whiten both the data and
the templates which span the space of physically allowable orbital period, orbital phase (epoch), and transit duration.  Although
the window sizes used to estimate the noise at the various time-frequency scales are large compared to the transit duration, 
the whitening process can lower the Signal-to-Noise Ratio due to the way that any signal present in the data perturbs the estimates of 
the noise (or whitening coefficients).  
The same is true
of any detrending of the data that is performed prior to a search.  For this reason, we do not detrend the data in any way
prior to the search other than to remove edge effects of spacecraft pointing at quarter and monthly boundaries 
where the spacecraft broke from data acquisition in order to beam the data back to Earth.  We also attempt to filter out coherent stellar oscillations by identifying and removing harmonics from each quarterly segment. This does, indeed, suppress such stellar variability, but also reduces the sensitivity of TPS to strong short period planetary signatures\citep{jchrist}. TPS normalizes each quarterly segment by its median and fills the monthly and quarterly gaps using methods that attempt to preserve the correlation structure of the observation noise and permit FFT-based approaches to be used in the subsequent search. 

After TPS has found a signal with a Multiple Event Statistic (MES) \citep{jmj2010} above threshold, however, it is 
possible to detrend and re-whiten the data and avoid any loss in Signal-to-Noise Ratio (SNR) that would otherwise 
occur because of the effect of the signal on the estimated trend and the whitening coefficient estimates.  Although 
this will not make the problem of detecting the signal any easier, since it can only be done after the initial detection, 
it does improve the discriminating power of the other statistical tests, or vetoes (discussed subsequently), that 
TPS subjects each candidate event to.  Prior to calculation of the veto statistics, the ephemeris of the candidate event is 
used to identify in-transit cadences (with some small amount of padding).  These in-transit cadences are then filled using an 
adaptive auto-regressive gap prediction algorithm.  A trend line is then estimated using a piecewise polynomial fitting algorithm 
that employs Akaike's Information Criterion to prevent over-fitting.  After removing this trend from the data, the 
whitening coefficients are then re-computed.  After removing the trend from the in-transit cadences, they are restored to the trend-removed data which are then whitened using the new whitening coefficients.  The potential candidate is then subjected to a suite of 
four statistical tests described below in section~\ref{vetoes}.   

\subsection{Search Templates and Template Spacing}
Mismatch between any true signal and the template used to filter the data degrades SNR.  This degradation due to 
signal-template mismatch can be decomposed into two separate types: shape mismatch, and timing mismatch.  The transit 
duration and assumed transit model affect the shape mismatch, while the template search grid spacing in orbital period and 
epoch affect the timing mismatch. 

TPS searches a total of 14 transit durations: 1.5, 2.0, 2.5, 3.0, 3.5, 4.5, 5.0, 6.0, 7.5, 9.0, 
10.5, 12.0, 12.5, and 15 hours.  These are logarithmically spaced, rounded to the nearest half hour (roughly the time per cadence), 
and augmented by a requirement to always search for 3, 6, and 12 hour pulses.  Prior to this run of the pipeline, TPS has simply used
a square-wave transit model. In \citet{seader2012}, it was shown through a Monte Carlo study, that with perfect duration and timing
match, the square wave, on average, mismatches a true signal by $~3.91$\%.  This translates directly into SNR loss.  This same 
Monte Carlo study was used to compute the integral average of all astrophysical models based on the  Mandel and 
Agol geometric transit model \citep{mandel} with limb darkening of Claret \citep{claret}, over the parameter space of interest \citep{seader2012}.  TPS now uses this averaged model to construct 
templates, which lowers the shape mismatch to only $~1.49$\%.  Lowering this shape mismatch also improves the sensitivity 
of the $\chi^2$ vetoes (discussed in the next section) which currently assume a perfect match between the signal and template.  In 
the next pipeline code release, the calculation of the $\chi^2$ vetoes will take into account the signal-template mismatch as 
described first in \citet{Allen} and later in \citet{seader2012}.  Using the new templates, the total shape mismatch (transit 
duration included) is only $~1.66$\% compared to $~4.32$\% for the square wave model.  

The mismatch in timing is a function of the template spacing in the period-epoch space.  This is controlled by a mismatch 
parameter which is essentially the Pearson Correlation Coefficient between two mismatched pulse trains 
\citep{jenkins1996,jmj2010}.  Previously, TPS 
required a match of $90$\%, which gave a timing mismatch alone of $~2.65$\% with the new astrophysically motivated templates for a 
total mismatch of $~4.31$\% including both shape and timing mismatches.  Since we are now running TPS on the NAS however, we can 
afford to search a finer grid of templates and have therefore tightened up the period-epoch match to $95$\%.  This parameter may be
increased further in the future but by increasing the number of search templates the false alarm rate also increases.  So there are 
tradeoffs to consider outside of just run time.  The false alarm vetoes keep the total number of false alarms at a manageable level
even with the increase to the period-epoch match.   

\subsection{False Alarm Vetoes}\label{vetoes}
The most substantial changes in terms of both impact on final results and extent in the codebase, are the changes to the 
false alarm vetoes since the last pipeline codebase release cycle.  During the transiting planet search, TPS steps through
potential candidates across period-epoch space with MES values exceeding the search threshold of $7.1\sigma$ in 
order of decreasing MES for each pulse duration.  The search continues until either TPS runs out of time on that pulse duration,
it hits the maximum allowable number of candidates to loop over (set to $1000$),
it exhausts the list without finding anything that passes all the vetoes, or it settles on something that passes all the vetoes. 
During the course of performing the search, TPS has the ability to remove up to two features in the data that contribute to 
detections that do not pass all the vetoes \citep{pt2013}.   After removing features, the 
period-epoch folding is re-done to generate a new list of candidates.

Candidate events are subjected to a suite of four statistical tests, two of which are new to this pipeline code release.  First, 
the distribution of the MES under the null hypothesis at the detected period is estimated by a Bootstrap test outlined in detail in 
Appendix~\ref{app:bootstrap} and in \citet{seaderBootstrap}.  From the estimated MES distribution, the threshold needed to 
achieve the false alarm probability equivalent to a $7.1\sigma$ threshold on a standard normal distribution ($~6.28 \times 10^{-13}$) is 
calculated by either interpolation or extrapolation.  If the whitener is doing its job perfectly, the MES 
distribution should be standard normal.   When extrapolation is needed, a linear extrapolation in log probability space 
is done which yields values that are typically conservative.  If any threshold values are suspect, or if the distribution can not be 
constructed for some reason, then this veto is not applied.  Otherwise, we require that the MES exceeds the calculated threshold
value to, in effect, ensure that we are making a detection that meets the false alarm criteria we have placed on the search.  The 
threshold used in this run for the bootstrap veto is too strict given that many real transiting planets are found in multiple planet 
systems.  The presence of transits for other candidate events, on a given target, in the null statistics will artificially elevate the threshold since they 
will add counts to MES bins above background.  In 
future runs, after appropriate tuning, the threshold will be relaxed to reduce the likelihood of rejecting real planetary signatures.

This Bootstrap veto does an excellent job at removing long period false alarms that are related to rolling band image artifacts.  
The image artifacts render the MES distribution non-white and potentially non-Gaussian after whitening.  
This means that $7.1\sigma$ is not as significant in comparison to the background at a given period for a given target 
than it would be in a purely standard normal distribution \textbf{so we must therefore require a higher threshold to achieve the desired false alarm rate}.  
Under the Neyman-Pearson criterion, which is employed in our detection strategy, the search threshold is 
determined from a required false alarm rate.  The Bootstrap veto allows us to map out a correction factor for each target and period
so that we can ensure the required false alarm rate is met.  This is sure to play an important role in upcoming planet occurrence 
rate studies where one must assess detectability of potential signals spanning the whole parameter space on every target.  The 
Bootstrap shows us that the MES distribution, and therefore detectability, is highly variable from target-to-target and across 
period space as well.

Candidates that pass the Bootstrap veto are then subjected to the Robust Statistic (RS) veto after the detrending and re-whitening 
described in Section~\ref{detrend} above.  This veto is described in detail in Appendix A of \citet{pt2013}.  The fit SNR is 
derived from robustly fitting the whitened data to a whitened model pulse train constructed using the ephemeris of the candidate 
event.  Previously, the threshold for this fit SNR was set to $6.4\sigma$ based on examination of the slope of the best fit line on
a plot of MES versus RS for KOIs detected with the correct ephemeris.  We found previously that $RS = 0.9 MES$, hence the 
$6.4\sigma$ threshold since we have a $7.1\sigma$ threshold on MES.  Now however, we find $RS = 1.15 MES$, which means we could 
in principle set the RS threshold at $~8.1\sigma$.  This makes sense considering that the whitening process lowers SNR and the MES has not been re-computed with the new whitener as the RS has.  For this run, we conservatively raised the threshold on the RS to $6.8\sigma$.  
There is an additional criterion applied during the RS test that affects only candidates with the minimum allowed 
number of transits (three transits).  We require that each transit has no more than $50$\% of its cadences with data quality weights 
less than unity \citep{pt2014}.  

If the RS threshold is exceeded, the candidate event is subjected to two separate \textbf{statistical} $\chi^2$ tests.  The first of these 
has been previously used and is described in \citet{pt2013} as $\chi^2_{(2)}$ with modifications as discussed in 
\citet{pt2014} and \citet{seader2012}.  This test breaks up the MES into different components, one for each transit 
event, and compares what is expected from each transit to what is actually obtained in the data assuming that there is indeed a 
transiting planet.  The current formulation assumes there is no mismatch between the signal and template.  In the next pipeline 
release however, the mismatch will be explicitly accounted for by modifying the way the number of degrees of freedom are calculated
as discussed in \citet{Allen} and \citet{seader2012}.  Since \citet{pt2014}, the use of $\chi^2_{(1)}$ \textbf{has} been dropped due to the
deficiencies discussed in \citet{seader2012}.  We have also dropped the use of $\chi^2_{(3)}$ as defined in \citet{seader2012}.  
We have however implemented a new $\chi^2$ veto, which tests the goodness of fit, that was presented in \citet{baggio} and later 
discussed in \citet{Allen}.  The new veto is dubbed $\chi^2_{(GOF)}$ and details of its construction are presented in 
Appendix~\ref{app:chisquare}.  The thresholds used for both $\chi^2$ vetoes are $7.0\sigma$.

The number of targets with a MES above threshold was 126,153, or ~63.5\% of all the targets.  The vetoes were then applied to this 
set of targets.  The bootstrap veto rejected 107,846 targets, the RS rejected an additional 3,553 targets, and the $\chi^2$ vetoes 
rejected an additional 2,056 targets. Note that there is a lot of overlap across the different vetoes for targets that were rejected, 
for example most of the targets vetoed by the bootstrap would also be vetoed by the $\chi^2$ vetoes, and targets vetoed by one 
version of the $\chi^2$ veto would also be vetoed by the other.  Together, this is a powerful set of false alarm vetoes each
with a very firm theoretical basis, that complement each other well to discriminate against the myriad of potential types of noise
that can masquerade as transiting planet signals.  The vetoes are not perfect however, and do prevent the generation of TCEs for some number of legitimate transiting planets.  We are in the process of fully characterizing their performance through transit injection studies.

\subsection{Detection of Multiple Planet Systems}
For the 12,669 target stars which were found to contain a threshold crossing event, additional TPS searches
were used to identify target stars which host multiple planet systems.  The process is described in \citet{hw2010} and in 
\citet{pt2013}. The multiple planet search incorporates a configurable upper limit on the number of TCEs per target, 
which is currently set to 10. This limit is incorporated for two reasons. First, limitations on available computing resources
translate to limits on the number of searches which can be accommodated, and also on the number of post-TPS
tests which can be accommodated. Second, applying a limit to the number of TCEs per target prevents a failure mode
in which a flux time series is so pathological that the search process becomes ``stuck,'' returning an effectively infinite
number of nominally-independent detections. The selected limit of 10 TCEs is
based on experience: to date, the maximum number of KOIs on a single target star is 7, which indicates that at this time,
limiting the process to 10 TCEs per target is not sacrificing any potential KOIs.

\begin{figure}
\plotone{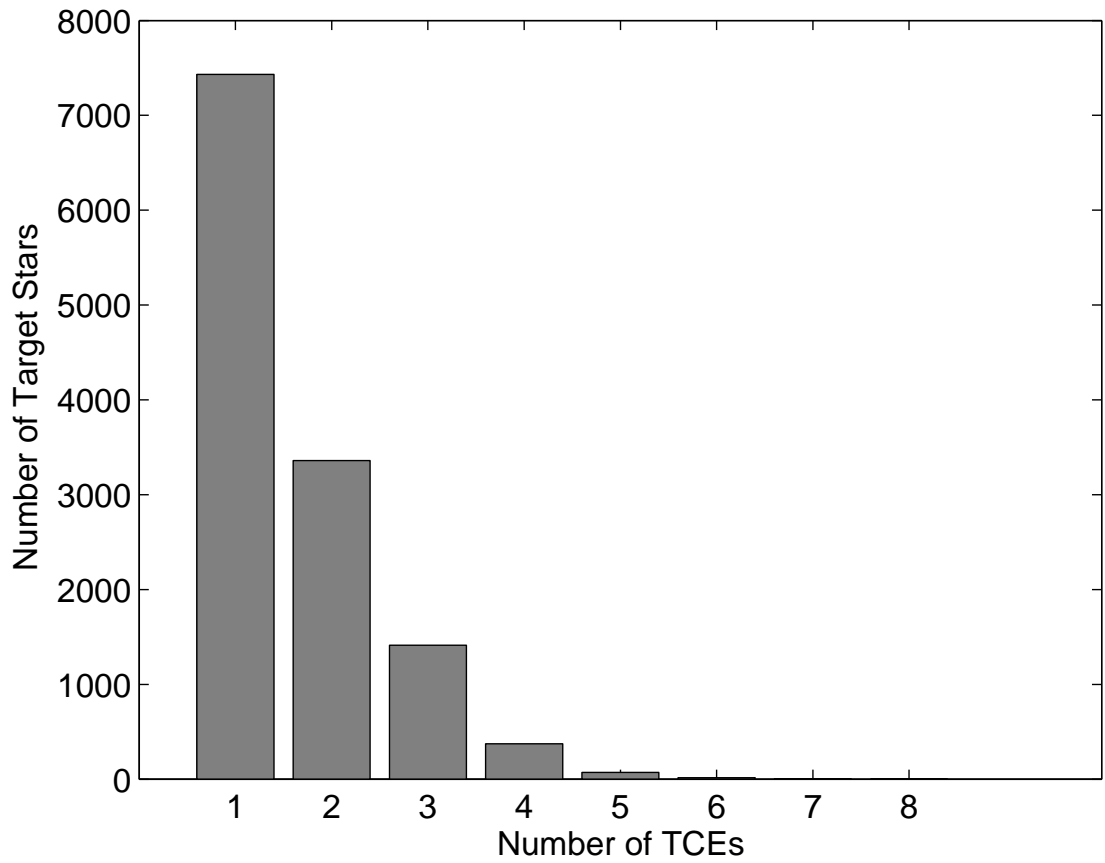}
\caption{Histogram of the number of targets having a particular number of TCEs.
\label{f2}}
\end{figure}

The additional searches performed for detection of multiple planet systems yielded 7,698 additional TCEs across 5,238 target stars, 
for a grand total of 20,370 TCEs. Figure \ref{f2} shows a histogram of the number of targets with each of the allowed numbers of
TCEs.  In this run, 16 targets produced 6 TCEs, 3 targets produced 7 TCEs, and 1 target had 8 TCEs.  
 Note that all of these TCEs are subjected to the full TPS process of detection and
vetoing described above.

In the analyses below, 3 targets that produced TCEs are not included.  This is due to the desire to limit
the analysis presented here to TCEs for which there is a full analysis available from the DV pipeline
module \citep{hw2010}.  These 3 targets failed to complete their DV analyses due to timing out and are thus
excluded. The Kepler Input Catalog (KIC) numbers for these 3 targets are: 5513861, 8019043, 10095469. The TCEs found
around KIC targets 5513861 and 10095469 were both short period (~0.676 and ~0.755 days respectively) and were found previously
and are contained in the TCE catalogs at the NASA Exoplanet Archive.  They were not made into KOIs by the TCE Review Team (TCERT).
The other TCE found on KIC target 8019043 has been found previously and is contained in the KOI catalog at the NASA Exoplanet
Archive and is labeled as being a false positive (KOI 6048.01). The 20,367 TCEs included
in this analysis \textbf{have been} exported to the tables maintained by the NASA Exoplanet 
Archive\footnote{http://exoplanetarchive.ipac.caltech.edu.}.

\section{Detected Signals of Potential Transiting Planets}\label{detected-signals}

\begin{figure}
\plotone{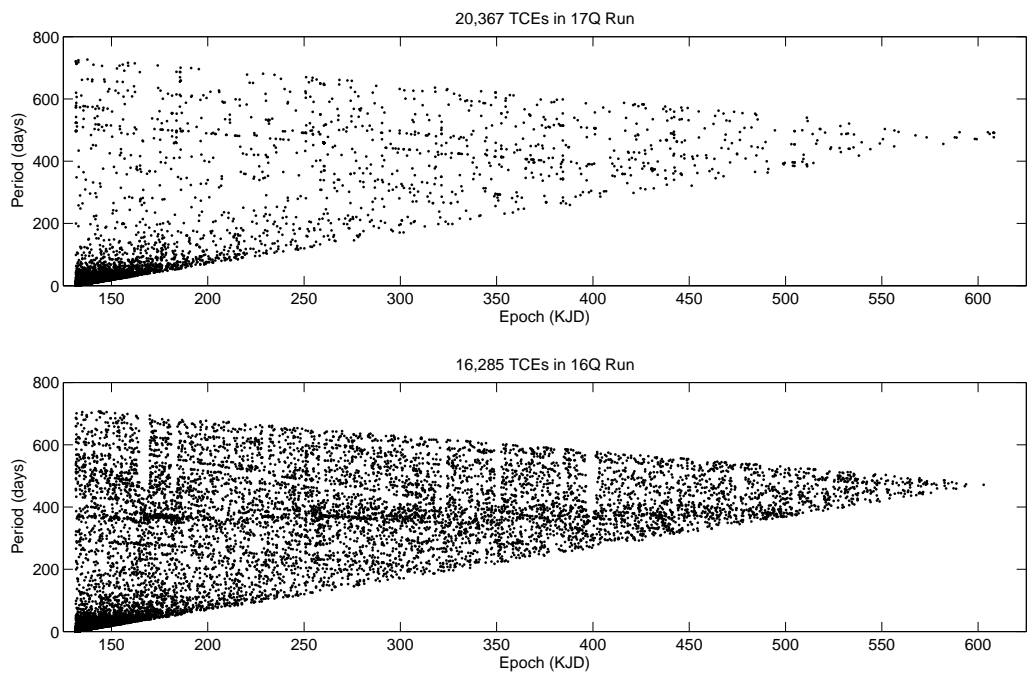}
\caption{Orbital period versus epoch for the 20,367 TCEs detected in Q1-Q17 of \kepler{} data (top); and for the 
16,285 TCEs detected in Q1-Q16 of \kepler{} data (bottom) as reported in \citet{pt2014}. Periods are in days, epochs are in 
Kepler-modified Julian Data (KJD), see text for definition.
\label{f3}}
\end{figure}

As described above, a total of 12,669 targets in the \kepler{} dataset produced TCEs in this run.  
For 7,431 of these targets, only one TCE was detected; for 5,238 targets, the multiple planet search detected additional
TCEs. The total number of TCEs detected across all targets was 20,367.
Figure \ref{f3} (top panel) shows the period and epoch of each of the 20,367 TCEs, with period in days and epoch
in Kepler-Modified Julian Date (KJD), which is Julian Date - 2454833.0, the latter offset corresponding to January 1, 2009,
which was the year of \kepler{}'s launch. Figure \ref{f3} also shows the same plot for the 16,285 TCEs detected in the 16
 quarter \kepler{}
dataset, as reported in \citet{pt2014}.  The axis scaling is identical for the two subplots, as is the marker size.  Several features
are apparent in this comparison.  First, the number of TCEs is larger along with the number of targets producing TCEs.
Second, as expected, the small addition of Q17 furnishes an 
increased parameter space available for detections, as shown by the upward and rightward expansion of the ``wedge'' in 
Figure \ref{f3} (bottom panel) from the Q1-Q16 to the Q1-Q17 results.
Third, the distribution of TCE periods appears to be much different in the current analysis compared to the analysis done in \citet{pt2014} with a drastic reduction of TCEs having long periods and large increase of short period TCEs.

\begin{figure}
\plotone{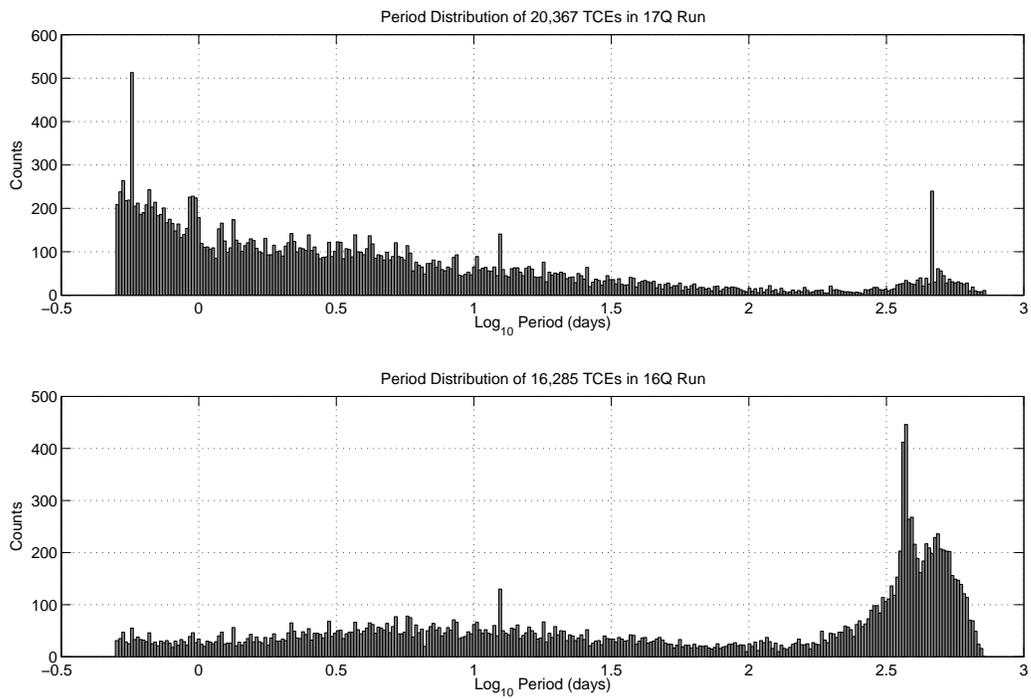}
\caption{Distribution of TCE periods plotted logarithmically. Top: 20,367 TCEs detected in Q1-Q17 of \kepler{} data; bottom: 
16,285 TCEs detected in Q1-Q16 of \kepler{} data as reported in \citet{pt2014}.
\label{f4}}
\end{figure}

The drastic change in the distribution of TCE periods can be seen more clearly in Figure \ref{f4}, which shows the distribution
of TCE periods on a logarithmic scale, with the Q1-Q17 results shown in the upper panel of the figure and the Q1-Q16
results in the lower panel.  The more recent search sharply reduces the number of long-period detections while simultaneously
recovering many of the short period detections which were wrongfully vetoed in the Q1-Q16 analysis presented in \citet{pt2014}.
The large reduction of long-period detections was largely due to the implementation of the bootstrap veto described in 
Section~\ref{vetoes}.  A majority of these long-period detections seen in analyses presented in \citet{pt2013} and \citet{pt2014},
were due to the rolling band image artifacts previously discussed as well as other issues.  
The bootstrap veto works well to veto these spurious detections
because there is an overabundance of non-Gaussian noise at long periods on these targets which requires a higher threshold for detection.  The 
shorter period detections were vetoed previously by the $\chi^2_{(3)}$ veto in \citet{pt2014} 
which was later determined to be flawed from a theoretical perspective.  Removal of this veto has led to the 
recovery of these shorter period detections.

Closer examination reveals that there are several peaks in the period histogram for the Q1-Q17 run that are highly localized given 
the fine binning.  The long period peak that persists at $~460$ days is due to the alignment of several separate data gaps that 
are relatively long.  There is currently an issue with the algorithm that fills these long data gaps.  Due to poor filling inside the
gaps, there can be features that cause ringing well outside the gapped region in the wavelet domain where the detections are made.  
We are in the process of fixing the algorithm that fills long gaps to prevent this issue in future releases.  Much of what is left 
at long periods outside of this strong peak is due mostly to uncorrected attitude tweak discontinuities, uncorrected sudden pixel 
sensitivity dropouts (due to cosmic rays or other energetic particles impinging on the CCD's pixels), argabrightening events, 
uncorrected positive-going outliers such as flares, momentum dumps, etc.  So these are largely just isolated events that produce a 
single significant Single Event Statistic (SES) which is then combined with noise typically to produce long period detections of 
only a few transits.   
 
At the extreme long periods, the $\chi^2$ vetoes have a low number of degrees of freedom and therefore 
often times don't have much discriminatory power to veto occurrences like this.  At shorter periods, these isolated features are 
either combined with enough noise to lower their MES sufficiently below threshold, or else the other vetoes remove them.  Other peaks
at shorter periods, such as the peak at $~12.45$ days and that at $~0.56$ days, are due to contamination from very bright sources such as V380 Cyg and RR Lyrae respectively \citep{jcough}.     

\begin{figure}
\plotone{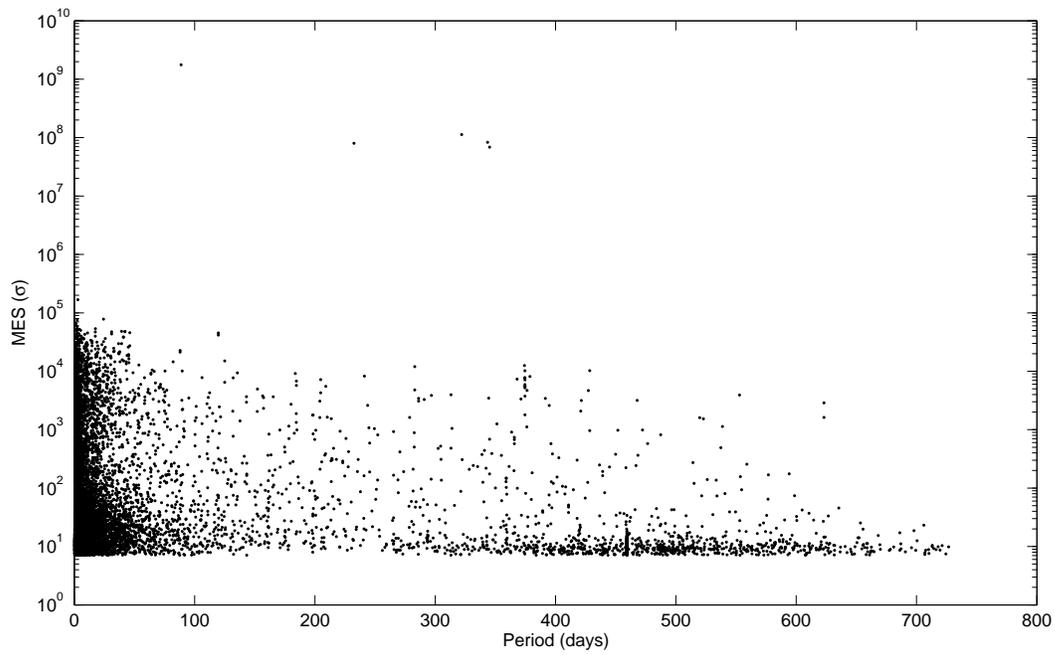}
\caption{Orbital period versus multiple event statistic for the 20,367 TCEs.
\label{f5}}
\end{figure}

Figure \ref{f5} shows the multiple event statistic (MES) and period of the 20,367 TCEs. The cluster of events with periods above
200 days, with relatively low multiple event statistic, are believed to be another representation of the long-period false alarms
discussed above. The relatively narrow cluster at approximately 380 days that was due to the ``one \kepler{} year'' instrument artifact
discussed and presented in the analogous figure in \citet{pt2014} is all but gone here.  The long-period TCEs are, for the most part,
relatively low MES for reasons already discussed above.  Improving our gap filling algorithm for long gaps and correcting some of the
other systematics discussed above should enable us to suppress much of the remaining long period false positives and determine if
anything significant remains to be detected which was hitherto screened.

\begin{figure}
\plotone{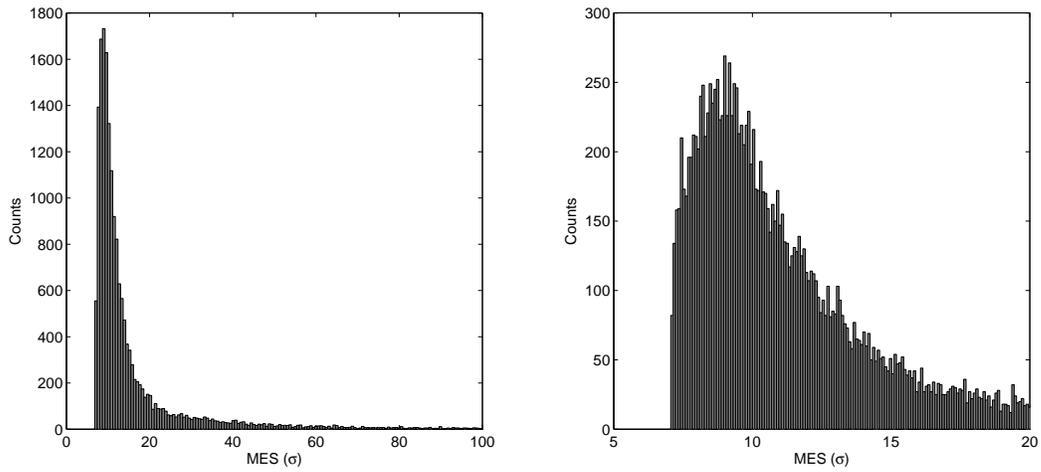}
\caption{Distribution of Multiple Event Statistics. Left: 17,785 TCEs with MES below 100$\sigma$. Right: 14,942 TCEs with MES below 
20$\sigma$.
\label{f6}}
\end{figure}

Figure \ref{f6} shows the distribution of
multiple event statistics: 17,785 TCEs with multiple event statistic below 100 $\sigma$ are represented in the left figure, while the
right hand figure shows the 14,942 TCEs with multiple event statistic below 20 $\sigma$. There were 2,582 TCEs with MES above 100
$\sigma$.  The bi-modality observed below 20 $\sigma$ in \citet{pt2014} has been removed.  The distribution now resembles quite well
the Extreme Value Distribution, as it should.  The mode of the distribution is $~9\sigma$.  

\begin{figure}
\plotone{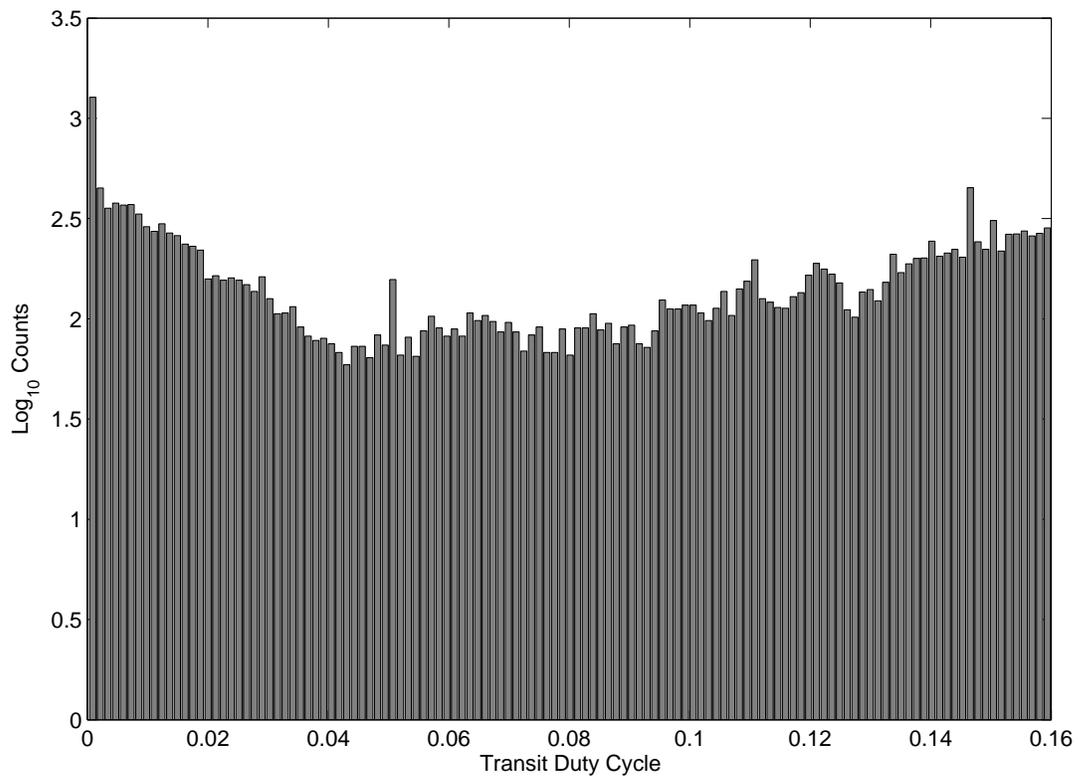}
\caption{Transit duty cycles of TCEs.
\label{f7}}
\end{figure}

Figure \ref{f7} shows a histogram of transit duty cycles of the TCEs.  
The transit duty cycle is defined to be the ratio of the trial transit pulse duration to the detected period of the TCE 
(effectively the fraction of time during which the TCE is in transit).  The overabundance of TCEs 
with extremely low duty cycles reported in \citet{pt2014} is significantly reduced whereas the re-introduction of the short period
TCE's previously vetoed (discussed above) is evidenced by a ramp up from a transit duty cycle of 0.05 up to 0.16.

\subsection{Comparison with Known Kepler Objects of Interest (KOIs)}\label{koi-comparison}

In what follows, comparisons are made between the TCEs returned by this latest pipeline run and the cumulative KOI catalog available at the NASA Exoplanet Archive\footnote{
https://exoplanetarchive.ipac.caltech.edu/}.  This archive of KOIs is built from many previous works including \citet{borucki1}, \citet{borucki2},
\citet{batalha1}, \citet{burke1}, \citet{rowe2015}, and \citet{mullally}.

As in past analyses \citep{pt2012,pt2013,pt2014}, we have identified a subset of the Kepler Objects of Interest (KOIs) which we use as a
set of test subjects for the TPS run.  TPS does not receive any prior knowledge about detections on targets; therefore, the re-detection
of objects of interest which were previously detected and classified as valid planet candidates
 is a valuable test to guard against inadvertent introduction of significant flaws into the detection algorithm.

The list of Q1-Q12, and Q1-Q16 KOIs has been analyzed and a set of high-quality ``golden KOIs'' identified for comparison to the Q1-Q17 TCEs.  This subset
of the full KOI list is a representative cross-section of all KOIs in the parameters of transit depth, signal-to-noise, and period.  It builds upon the list
used in \citet{pt2014} but also now excludes those KOIs that are labeled as false positives at the NASA Exoplanet Archive.

\begin{figure}
\plotone{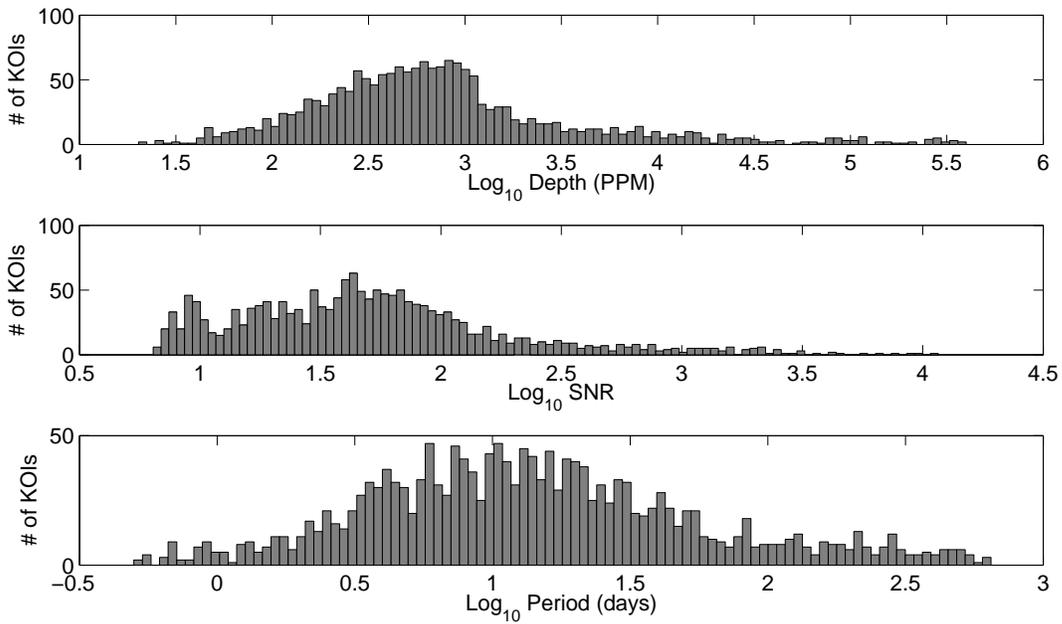}
\caption{Parameter distribution of ``golden KOIs'.''  Note the use of logarithmic horizontal axes.
\label{f8}}
\end{figure}

The ``golden KOI'' set includes 1,752 KOIs across 1,483 target stars.  Figure \ref{f8} shows the distribution of estimated transit depth,
signal-to-noise ratio, and period for the ``golden KOIs.''  Out of these, 1,411 target stars produced one or more TCEs, while 72 target stars
did not.  Of the 72 target stars which failed to produce a TCE, all but two produced one, and only one, KOI per target (total of 74 KOIs in all) in previous pipeline runs.  
Only 5 of the 74 KOIs had periods less than 10 days, so the issue described in \citet{pt2014} of a stellar harmonics removal algorithm removing short period transits is 
not a likely culprit here.   Out of the 74 missed KOIs, 48 ran through the search loop in TPS only a single time and so a precise determination of why 
they were missed can be made.  

For half of them, TPS failed to latch onto the correct period and therefore made no detection.  The other half were
vetoed evenly between the newly implemented bootstrap veto, and the $\chi^2$ vetoes.  These were all very close to threshold, so re-tuning the thresholds would
likely bring all these missed KOIs back.  As discussed in \citet{seader2012}, the $\chi^2$ vetoes assume a perfect match between the signal and template.  In the next pipeline
development cycle, this assumption will no longer be made and the thresholds used on the $\chi^2$ vetoes will be adjusted to allow for mismatch.  This adjustment will act to
lower the threshold, thereby making it easier to detect valid transit signals that are near threshold.  Injection studies so far have shown that this is very promising for 
flattening out the response of the vetoes across period space where previously there was a slight reduction in recoverability of injected signals at long periods.

\subsubsection{Matching of Golden KOI and TCE Ephemerides}

Detection of a TCE on a ``golden KOI'' target star is a necessary but not sufficient condition to conclude that TPS is functioning properly.  An
additional step is that the TCEs must be consistent with the expected signatures of the KOIs.  This is assessed by comparing the ephemerides
of the KOIs and their TCEs, as described in \citet{pt2013}; the ephemeris-matching process also implicitly compares the numbers of KOIs and
TCEs on each target star, which exposes cases in which, on a given star, some but not all KOIs were detected.

\textbf{There were 1,678 ``golden KOIs'' (the full 1,752 with the 74 that did not produce TCEs removed) on the 1,411 target stars which produced TCEs, and for 1,664 it was possible to find a match to a TCE.}  The fourteen KOIs which did
not produce TCEs were KOI 1101.01, KOI 492.02, KOI 2971.02, KOI 649.02, KOI 6178.02, KOI 6182.02, KOI 3088.02, KOI 6191.02, KOI 600.02, KOI 1310.02, 
KOI 351.02, KOI 351.03, KOI 351.04, and KOI 351.05 .  KOI 1101.01 is a short-period candidate (2.84 days) also missed in \citet{pt2014}, and was
most likely removed by the narrow-band oscillation algorithm; the second candidate on this target, with a period of 11.4 days, was detected with a correct
ephemeris match.  KOI 351, aka Kepler-90, is a multi-planet system with significant transit timing variations (TTVs) \citep{cabreraKepler90}; since TPS requires highly periodic
signals to produce a valid detection, its performance on this system has always been poor, and the 
bootstrap veto rejects the planets in this system affected by the large TTVs.  The remaining KOIs were seen at the correct ephemeris by TPS 
but were vetoed by either the bootstrap or $\chi^2$ tests.    

\begin{figure}
\plotone{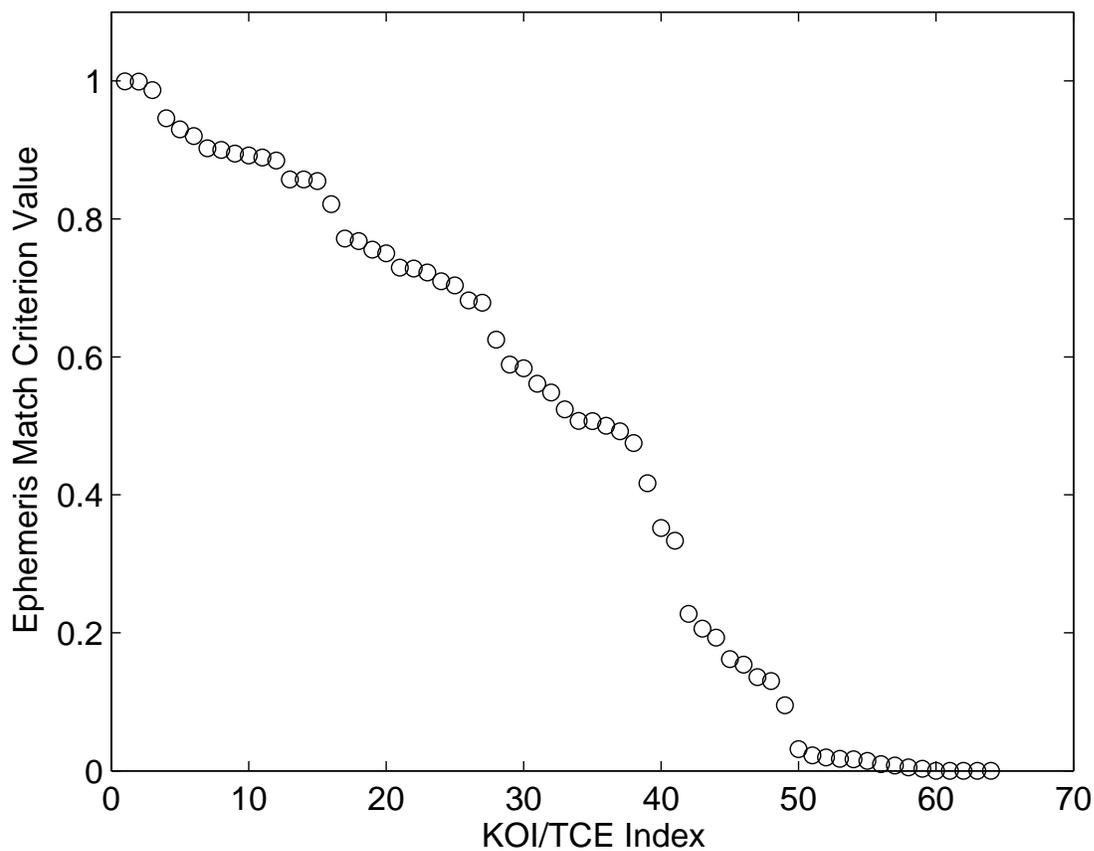}
\caption{Ephemeris match value for 64 ``golden KOI'' KOI-TCE matches with a criterion value less than 1, sorted into descending order.  A total of 1,600 KOI-TCE matches have an ephemeris match criterion value of 1.0.
\label{f9}}
\end{figure}

Figure \ref{f9} shows the value of the ephemeris match criterion for the 64 KOI-TCE matches where the value was less than 1, sorted into descending order.  A total of 1,600 
KOI-TCE matches have a criterion value of 1.0, indicating that each transit predicted by one ephemeris corresponds to a transit predicted by the
other, to within one transit duration.  In these cases, it has been assumed that TPS correctly detected the ``golden KOI'' in question and no further
analysis was performed.

In the 64 cases in which the ephemeris match was imperfect, each KOI-TCE match was manually inspected.  The disposition of the results is as
follows:
\begin{itemize}
\item In 33 cases, the TCE actually matches the KOI, but the value of the match parameter does not reflect this; in general this is because the 
KOI ephemeris was derived with early flight data, requiring extrapolation to determine the transit times late in the mission and permitting an
accumulation of error in the KOI transit timings relative to the actual timings
\item In 14 cases TPS detected the KOI at the correct period but the epoch was off by an integer multiple of periods
\item In 13 cases TPS failed to detect the KOI on the ``golden KOI'' list for a given target but did detect a different KOI on the same target.
\item In 1 case TPS detected the planet but at twice the orbital period of the KOI
\item In 1 case TPS detected the planet but at a third of the period of the KOI
\item In 2 cases TPS failed to detect the KOI on the ``golden KOI'' list but the detections can not easily be dismissed as 
being false positives without further study and scrutiny.  
\end{itemize}
In conclusion, out of the 64 KOI-TCE pairs which have imperfect ephemeris matches, 15 actually constitute a failure of the detection algorithm.
Table \ref{t1} lists the KOIs used in the ``golden KOI'' set, KOI and TCE ephemerides, and the corresponding ephemeris match parameter. 

\subsubsection{Matching of KOI and TCE Ephemerides}
The detailed analysis of the performance of the search on the ``golden KOI'' set is useful in identifying problems and gives us an idea for how well the
search should perform on the full set of KOIs.  Here we examine the actual performance on almost the entire cumulative set of KOIs available at the
NASA Exoplanet Archive.  Currently, there are a total of 7,305 KOIs, on a total of 6,150 targets, in the cumulative KOI table.  
The ``golden KOI'' set is a subset of this largest set of KOIs.  Since the purpose of this
analysis is to assess the performance of the search,  KOIs labeled as ``false positive'' or ``not dispositioned'' are removed from the set of KOIs.
This leaves a set of 4,173 KOIs across 3,196 targets.  It's important to note that we would not expect all of these to be detected since some have Transit 
Timing Variations (TTVs) that made them more easily detectable with a smaller amount of data than the full 17Q used for this run.  We make no effort to 
exclude these KOIs from the set (and in fact, even the ``golden KOIs'' analyzed above contain some KOIs that exhibit TTVs).

\begin{figure}
\plotone{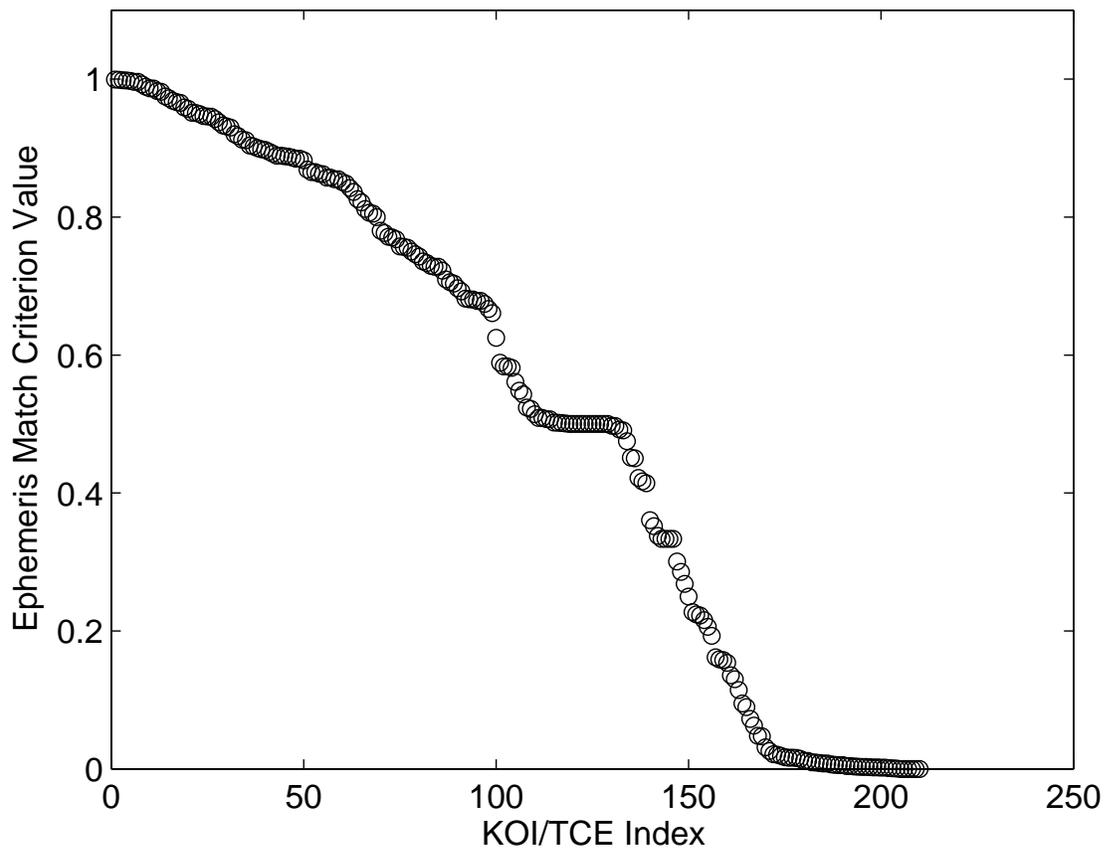}
\caption{Ephemeris match value for 210  KOI-TCE matches with a criterion value less than 1, sorted into descending order.  A total of 3,599 had an ephemeris match criterion value of 1.0.
\label{f10}}
\end{figure}

\begin{figure}
\plotone{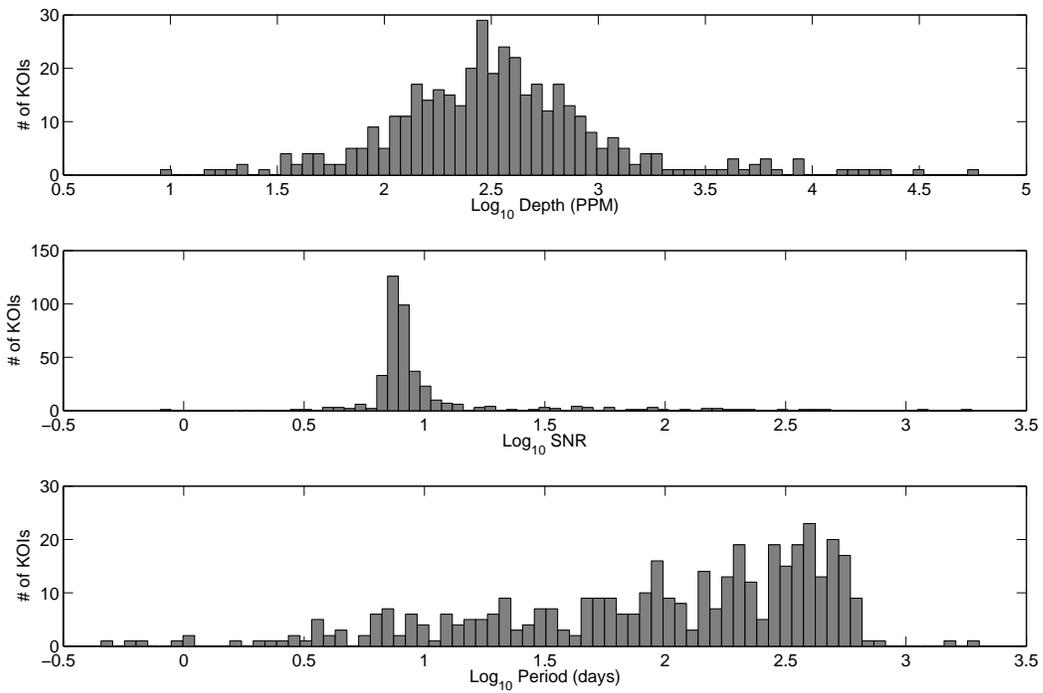}
\caption{Parameter distribution of KOIs that were not detected.  Note the use of logarithmic horizontal axes.
\label{f11}}
\end{figure}

Of the 4,173 KOIs in this set, it was possible to find matches for 3,809, leaving 364 unmatched to any TCE.  Of the 3,809 for which matches were found, 3,599
of them had an ephemeris match parameter of 1.0.  Figure \ref{f10} shows the value of the ephemeris match criterion for the 210 KOI-TCE matches with a value less than 1, 
sorted into descending order.  For the 210 with an ephemeris match less than 1:
\begin{itemize}
\item In 124 cases, the TCE actually matches the KOI, but the value of the match parameter does not reflect this; in general this is because the 
KOI ephemeris was derived with early flight data, requiring extrapolation to determine the transit times late in the mission and permitting an
accumulation of error in the KOI transit timings relative to the actual timings
\item In 21 cases TPS detected the KOI at the correct period but the epoch was off by an integer multiple of periods
\item In 23 cases TPS detected the planet but at a harmonic or sub-harmonic of the orbital period of the KOI
\item In 42 cases TPS failed to detect the KOI.
\end{itemize}
This means that TPS failed to detect 406 of the KOIs in this set.  Figure \ref{f11} shows the distribution of KOI depth,  KOI SNR,
  and KOI period (as given in the table at the NASA Exoplanet Archive).  The majority of these missed KOIs are low SNR events skewed toward 
long period.  A concerted effort is underway to tune the vetoes previously discussed, which should result in the recovery of these KOIs.

\begin{figure}
\plotone{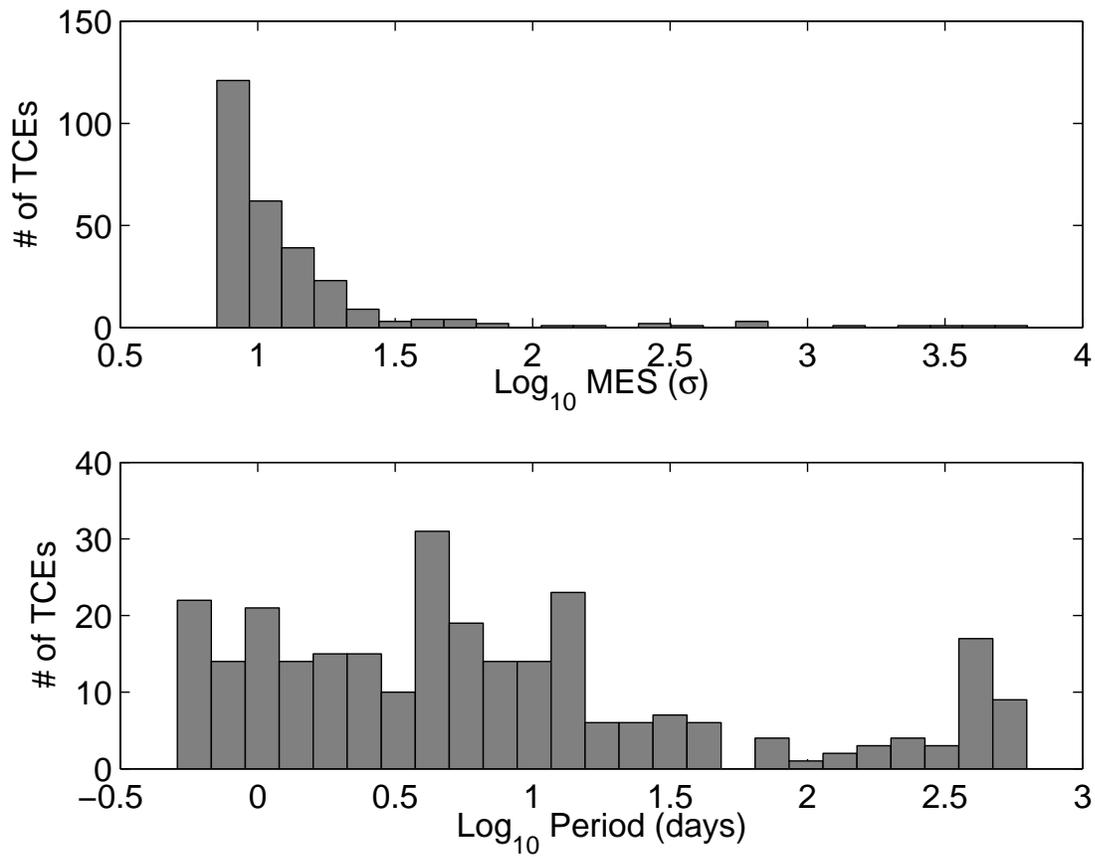}
\caption{Distribution of TCE MES and periods plotted logarithmically for 280 unmatched TCEs. 
\label{f12}}
\end{figure}

 On the 3,196 targets with KOIs in this set, there were 280 TCEs that did not match with a KOI (in this set).  Figure \ref{f12} shows the distribution of MES
and period of the TCEs.  These are primarily low MES events skewed toward short period.  Expanding the set of KOIs to include every KOI (no matter what
the disposition) shows that 51 of the 280 match a KOI labeled as a false positive or a KOI that has not yet been dispositioned. This leaves a total of 229
TCEs that do not match a KOI on the 3,196 targets. Note that some of these new TCE's may actually have been found in previous runs
but were excluded from becoming KOIs by the TCERT process.  No effort has been made to trim those out since this is a new data set
and a fresh look should be taken at them by TCERT.   These new TCE's (along with all the other TCEs)
 will be vetted by TCERT soon for dispositioning. Some of them may become new KOIs.  These 229 new TCEs will be briefly analyzed below in section \ref{dv}.

\subsubsection{Conclusion of TCE-KOI Comparison}
TPS recovered $~90.3$\% of the KOIs in the full set of vetted KOIs labeled as being planet candidates or confirmed candidates.  On the set of ``golden KOIs'',
TPS detected $~99.1$\%.  The missed KOIs were overlooked by TPS largely due to the overly aggressive thresholds imposed on the veto statistics\footnote{Note that many of these KOIs are marginal and have been shown through separate analyses to be unlikely planet candidates \citep{JenkinsNew} but they warrant additional scrutiny. So we do not expect all of them to be detected even if our detection algorithm were perfect.}.
We have successfully eliminated a majority of the long period false \textbf{alarms} and recovered a majority of the lost short period detections 
presented in \citet{pt2014}.  This is the first time the statistical bootstrap has been implemented and used as a veto inside of TPS and the
analysis of these results has served to illuminate the fact \textbf{that this} veto, and the others, need to be tuned through transit injection and studied
for other potential improvements.  We firmly believe a balance can be made whereby we can recover the missing KOIs presented here, but still reject the long
period false alarms that have been an issue in the past.  This, along with further mitigation of spurious long period false alarms, will constitute a bulk of 
remaining effort in TPS as the \kepler{} primary mission comes to a close.
The full list of 20,367 Q1-Q17 TCEs found and analyzed in this processing run \textbf{have been exported to, and will be} maintained by, the NASA Exoplanet Archive.

\section{Data Validation Results}\label{dv}
The Data Validation module attempts to validate each TCE returned by TPS by performing a set of tests.   A complete description of what DV does is
outside the scope of this paper, but the interested reader can take a look at \citet{hw2010} and \citet{jt2014}. The Data Validation module of the data 
processing pipeline has undergone significant improvements in this latest release cycle.  To mention some: 
ephemeris matching against known KOIs and confirmed planets, completely reworked statistical bootstrap(same as that described for TPS), 
improvements to the test for weak secondary events, difference image generation in quarters where all observed transits are 
overlapped by transits of other candidates, and numerous updates to the DV full report and one-page summaries that are produced for every TCE.  Here
we present some of the results produced by DV for all the TCEs and also take a closer look at the 229 new TCEs, found on the large set of KOI targets 
examined above, that did not match an existing KOI. 

\subsection{Aggregated Results}  
\begin{figure}
\plotone{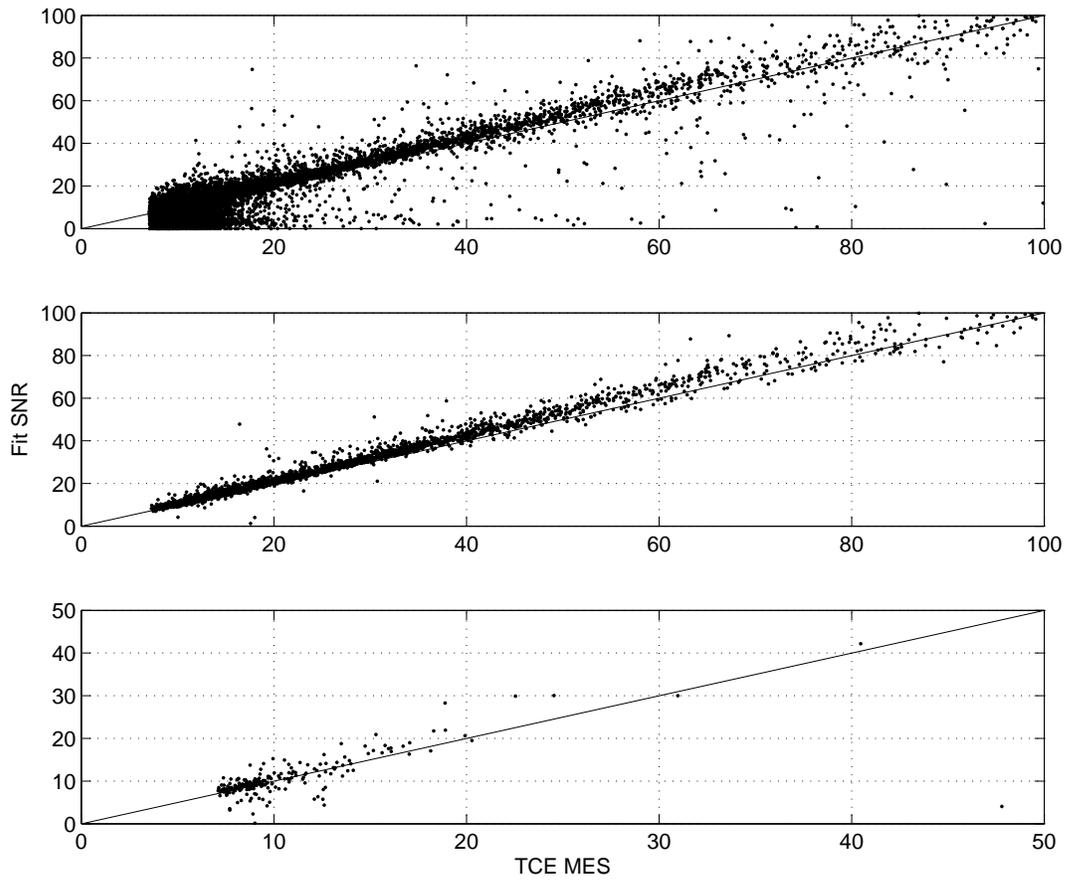}
\caption{DV fit SNR vs. TCE MES. Top: All 20,367 TCEs.  Middle: 3,599 TCEs with ephemeris match of 1 to a KOI. Bottom: 229 new TCEs on 
target stars with KOIs.
\label{f13}}
\end{figure}

During the Data Validation step, a Levenberg-Marquardt algorithm is employed to search for the best astrophysical model across the parameter space of
orbital period, transit duration, impact parameter, ratio of planet radius to host star radius, and the ratio of orbital semi-major axis to the host star radius.
The astrophysical models are constructed using the geometric transit model of Mandel and Agol \citep{mandel} with limb darkening of Claret \citep{claret2011}.
The SNR of this fit should, in general, be slightly larger than the MES that TPS returns since the ephemeris is more refined, the signal-template mismatch
should be lower(due to both barycentric time correction and the use of actual astrophysical models for the target star), 
and the whitener should not be degrading signal since it's recomputed for in-transit cadences after removing their effect. A fit SNR that is 
significantly different from the MES could indicate that the TCE was produced by some phenomenon other than a transiting planet. Figure \ref{f13} 
shows how the SNR from the fit compares to the MES for all 20,367 TCEs, 3,599 TCEs that match existing KOIs with an ephemeris match of 1, and the 229 new TCEs.
Note that the 1,059 TCEs that failed to produce valid fits have been excluded and the axes have been restricted to focus on the bulk of the population.

\begin{figure}
\plotone{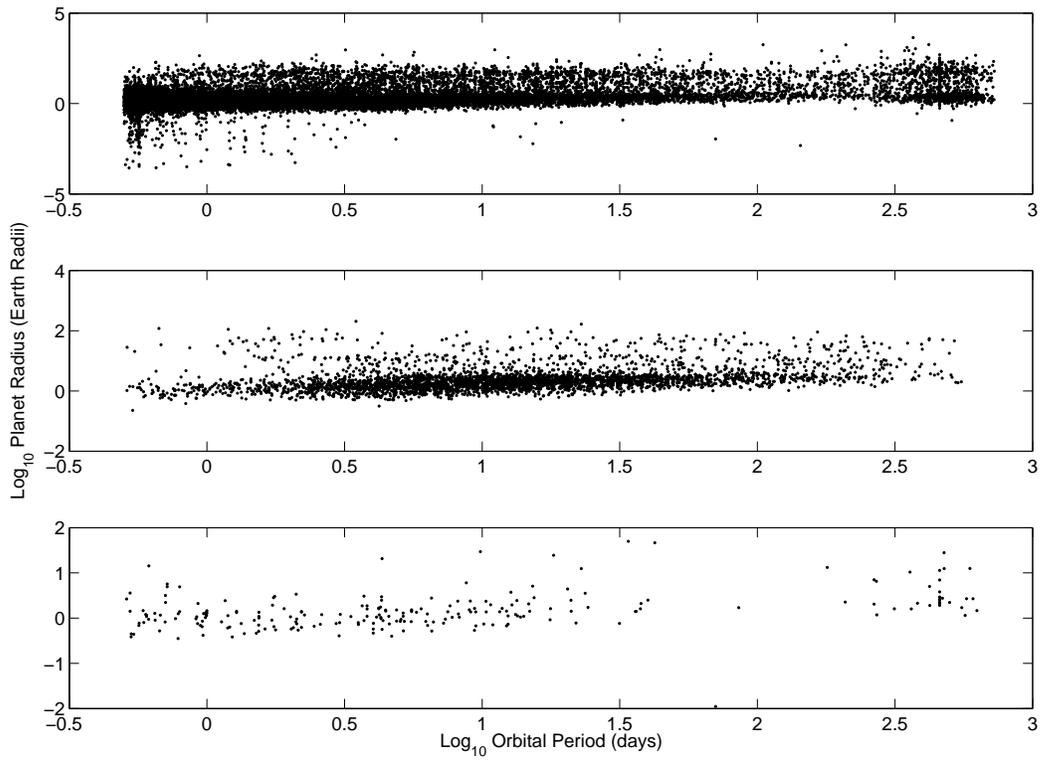}
\caption{Planet radius vs. orbital period on a logarithmic scale. Top: All 20,367 TCEs.  Middle: 3,599 TCEs with ephemeris match of 
1 to a KOI. Bottom: 229 new TCEs on target stars with KOIs.
\label{f14}}
\end{figure}

Using updated KIC parameters \citep{huber} and the results of the DV fitting process, the planet radius can be determined.  Figure \ref{f14} shows planet radius 
versus orbital period for the same groups of TCEs.  Clearly visible is the set of spurious long period TCEs as well a set of TCE's with both very large 
and very small planet radii which are clearly unphysical.

\begin{figure}
\plotone{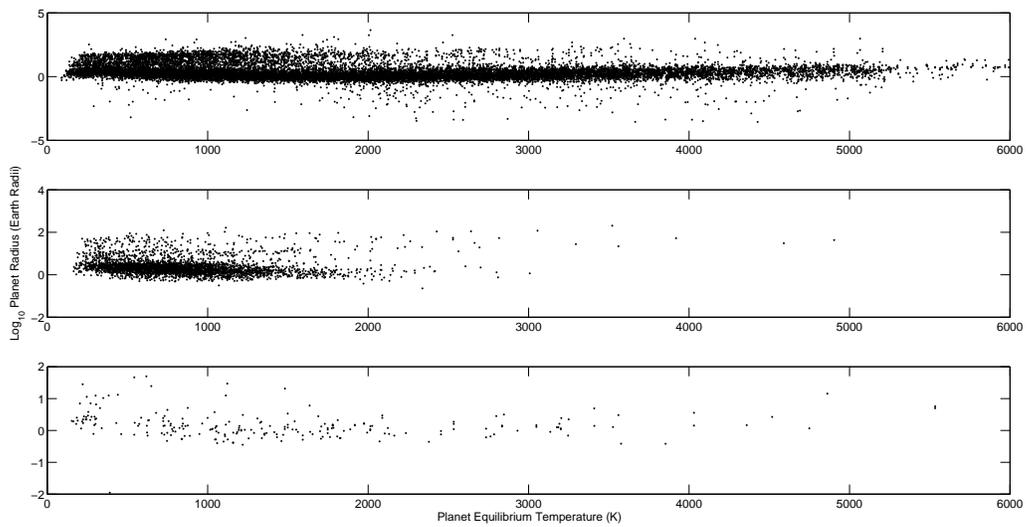}
\caption{Planet radius (on a logarithmic scale)vs. planet equilibrium temperature.  Top: All 20,367 TCEs.  Middle: 3,599 TCEs with ephemeris match of 
1 to a KOI. Bottom: 229 new TCEs on target stars with KOIs.
\label{f15}}
\end{figure}

Assuming perfect redistribution of heat and an albedo of 0.3, the planet equilibrium temperature can be calculated from
\begin{equation}
T_{eq} = T_{*} (1-\alpha)^{1/4} \sqrt{{R_{*}} \over {2 a}},  
\end{equation}
where $T_{*}$ is the effective temperature of the host star, $\alpha$ is the albedo, $R_{*}$ is the radius of the host star, and $a$ is the semimajor axis of the planet's orbit.
  Figure \ref{f15} shows a plot of the planet radius versus this equilibrium temperature for the same groups of TCEs.

\subsection{New Threshold Crossing Events}
Here we report on eight of the new TCEs that have never been found in a prior run, six of which are interesting from the standpoint of habitability (near 
Earth size, planet effective temperature near Earth's, and fairly significant SNR), and two of which are interesting because they are sub-Earth size. 
\textbf{It should be emphasized here that these are merely new TCEs and there is much work to be done on each of these in order for them to be promoted as planetary candidates.}
The one-page summaries produced by DV are given here but are also available at the NASA Exoplanet Archive, along with their full DV report counterparts that contain
much more information about each.
   
\begin{figure}
\plotone{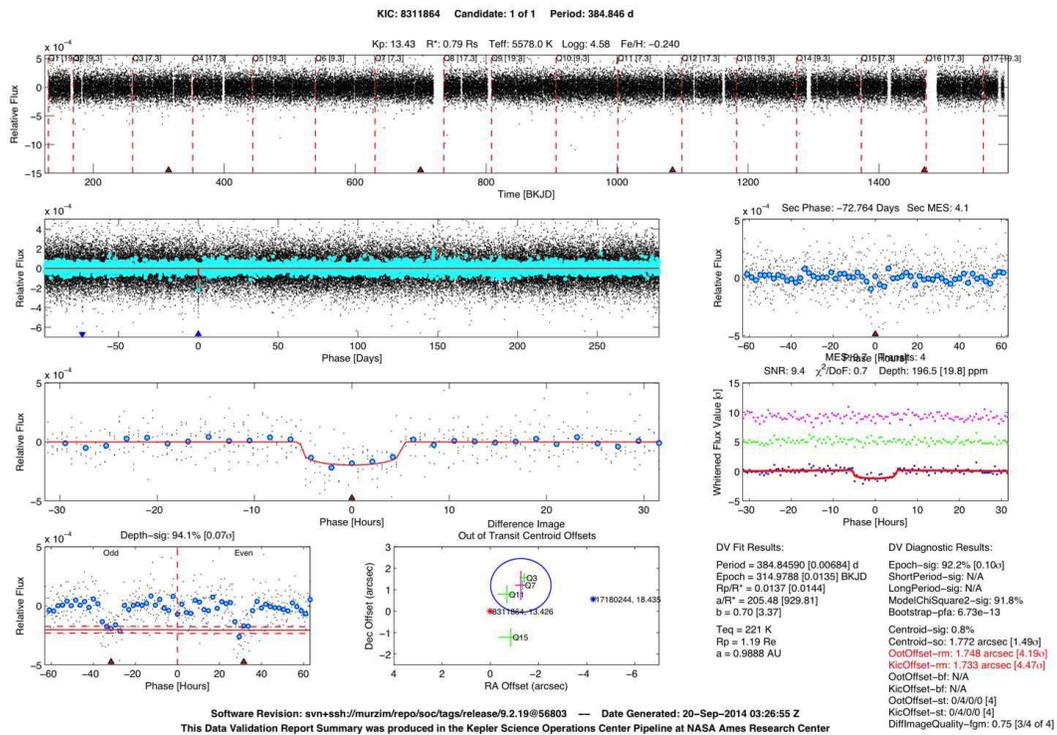}
\caption{One-page DV summary for a new candidate on KIC target 8311864.
\label{f16}}
\end{figure}

\begin{figure}
\plotone{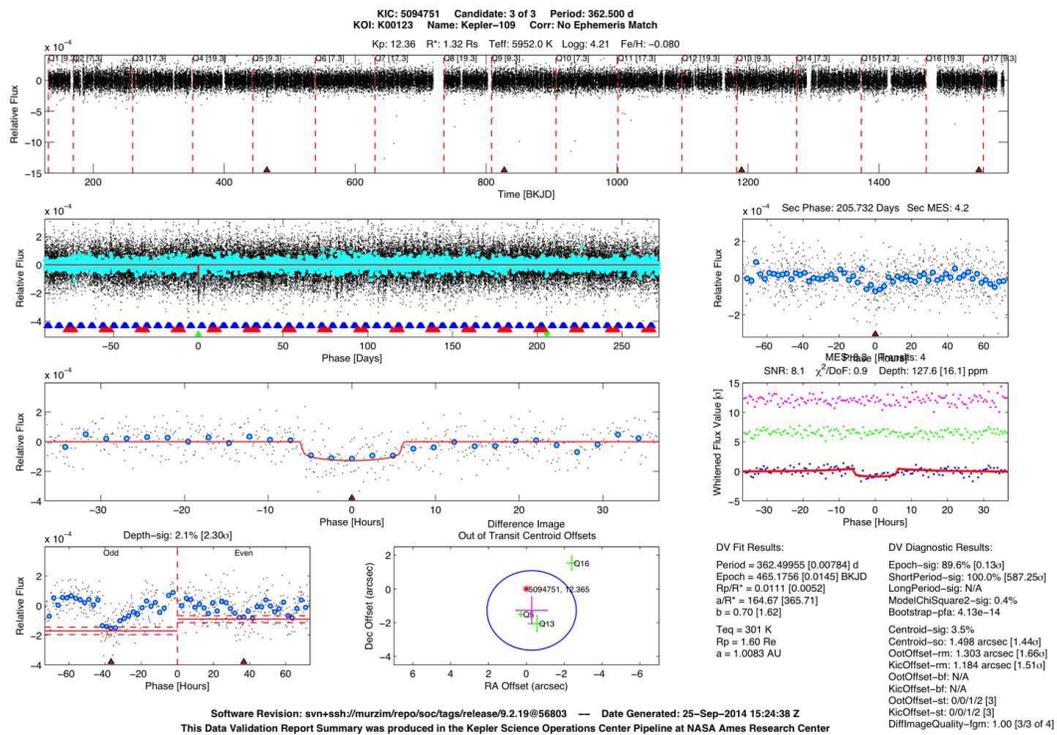}
\caption{One-page DV summary for a new candidate on KIC target 5094751.
\label{f17}}
\end{figure}

\begin{figure}
\plotone{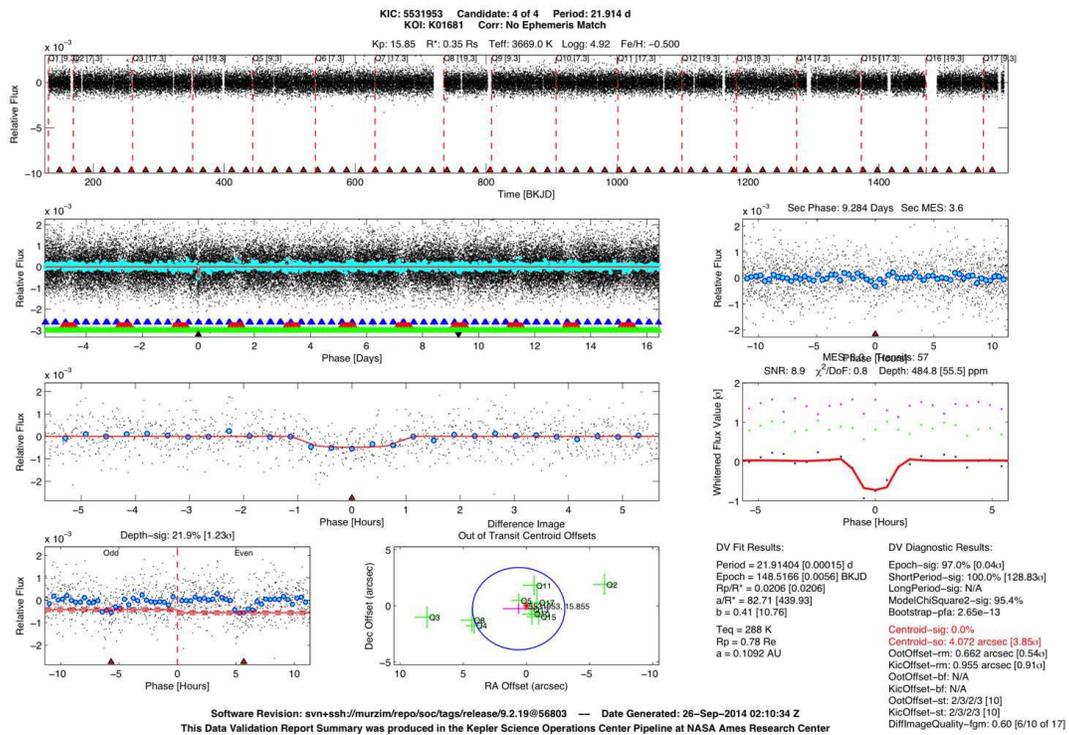}
\caption{One-page DV summary for a new candidate on KIC target 5531953.
\label{f18}}
\end{figure}

\begin{figure}
\plotone{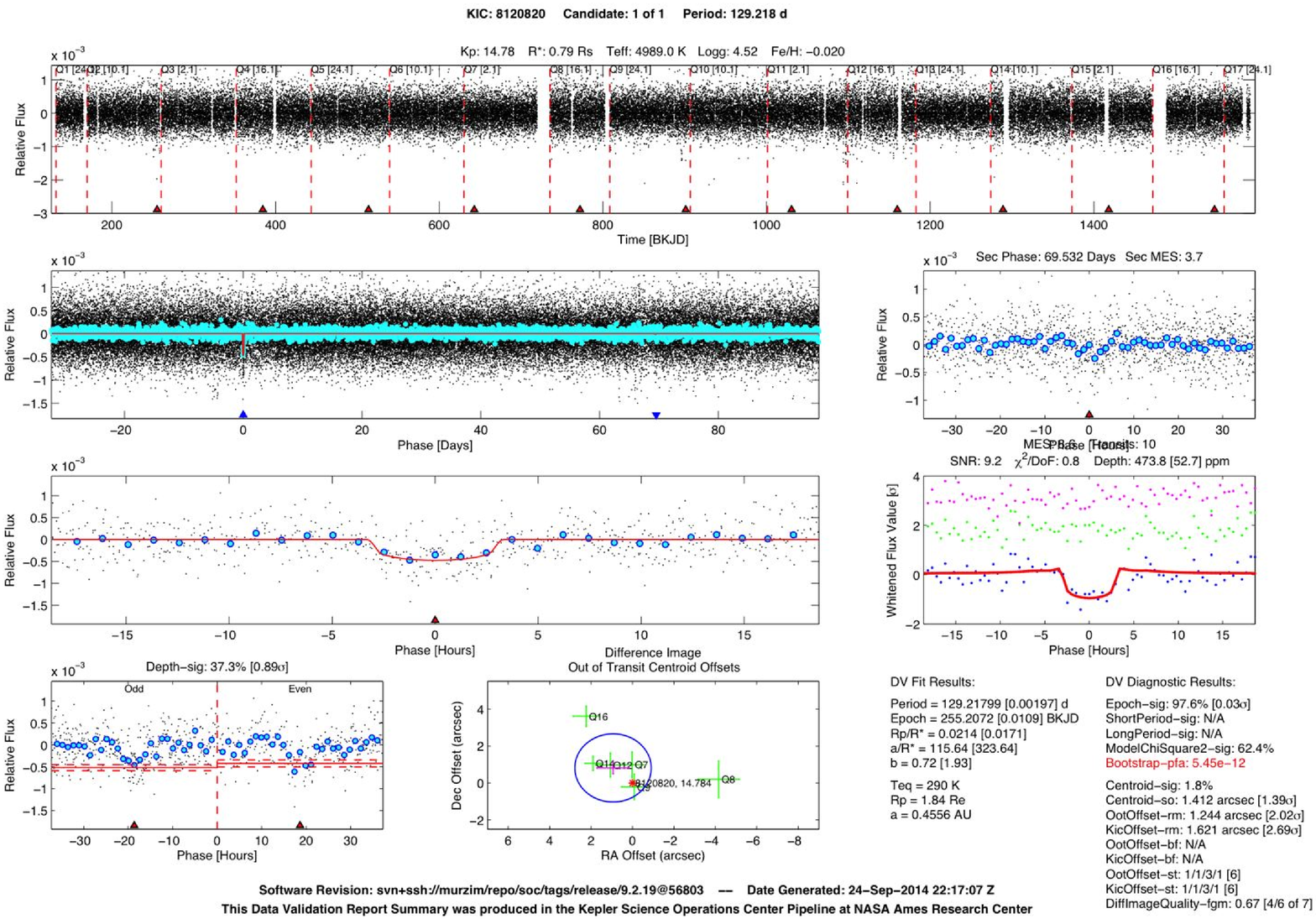}
\caption{One-page DV summary for a new candidate on KIC target 8120820.
\label{f19}}
\end{figure}

\begin{figure}
\plotone{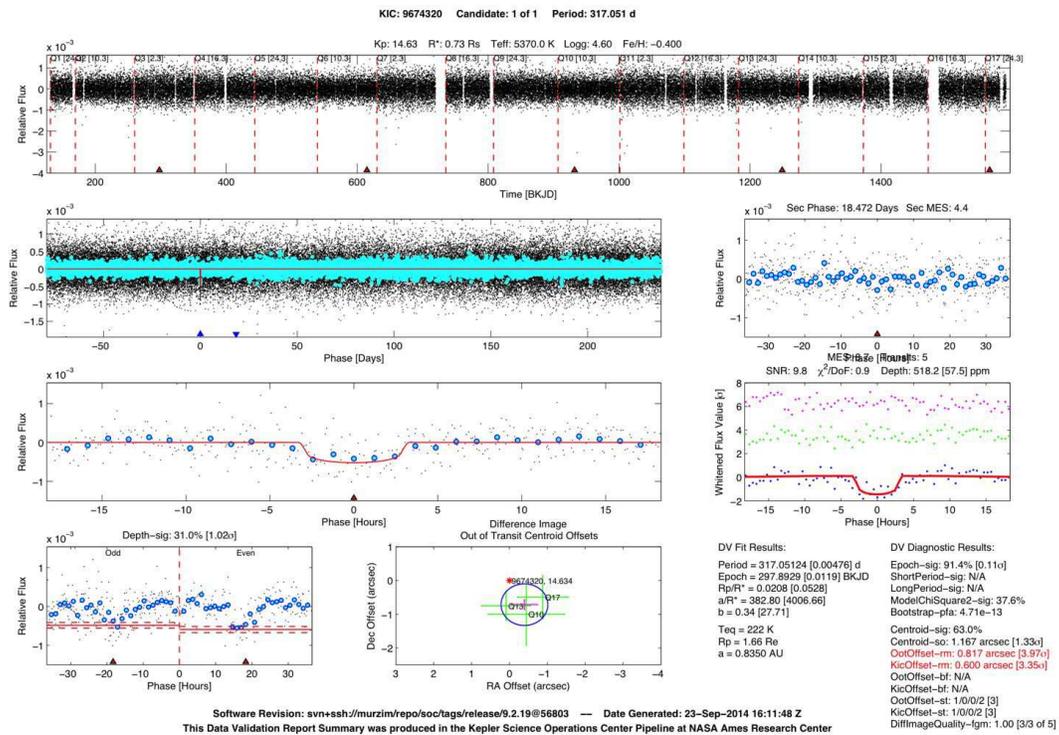}
\caption{One-page DV summary for a new candidate on KIC target 9674320.
\label{f20}}
\end{figure}

\begin{figure}
\plotone{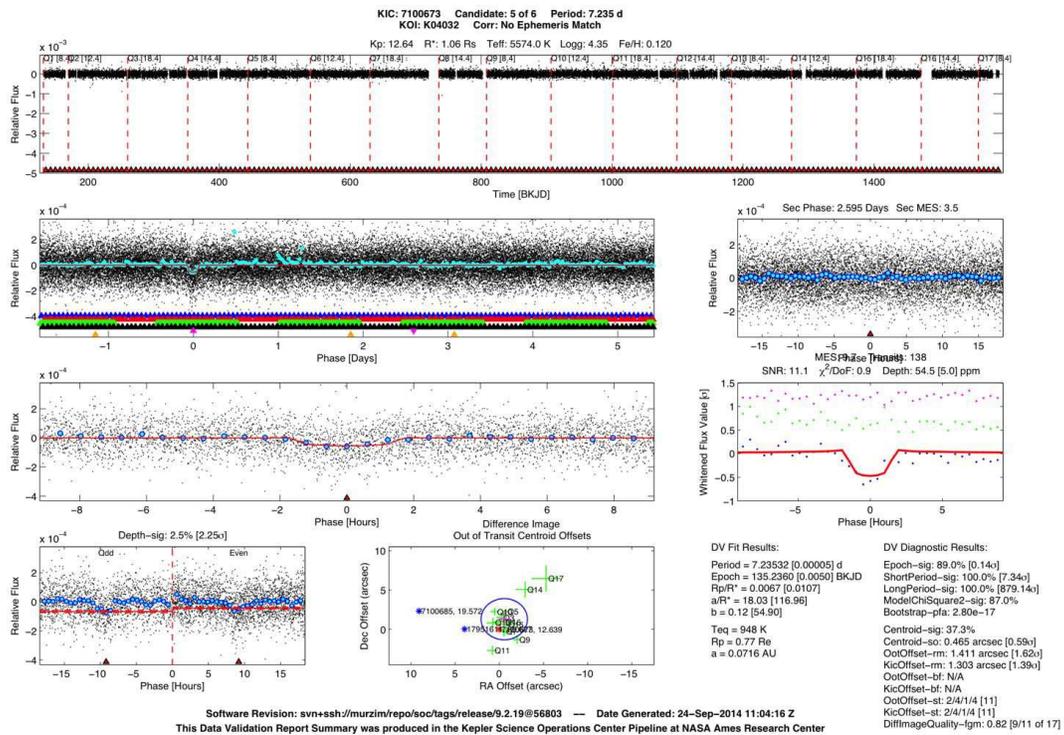}
\caption{One-page DV summary for a new candidate on KIC target 7100673.
\label{f21}}
\end{figure}

\begin{figure}
\plotone{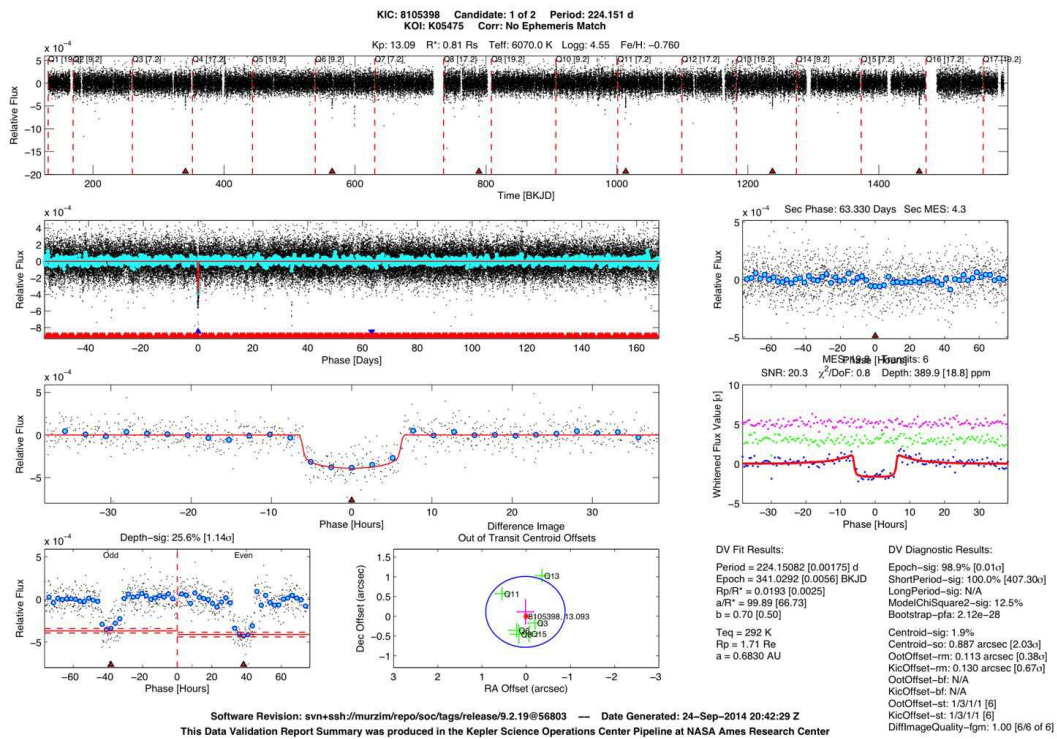}
\caption{One-page DV summary for KOI 5475.01 on KIC target 8105398.
\label{f22}}
\end{figure}

\begin{figure}
\plotone{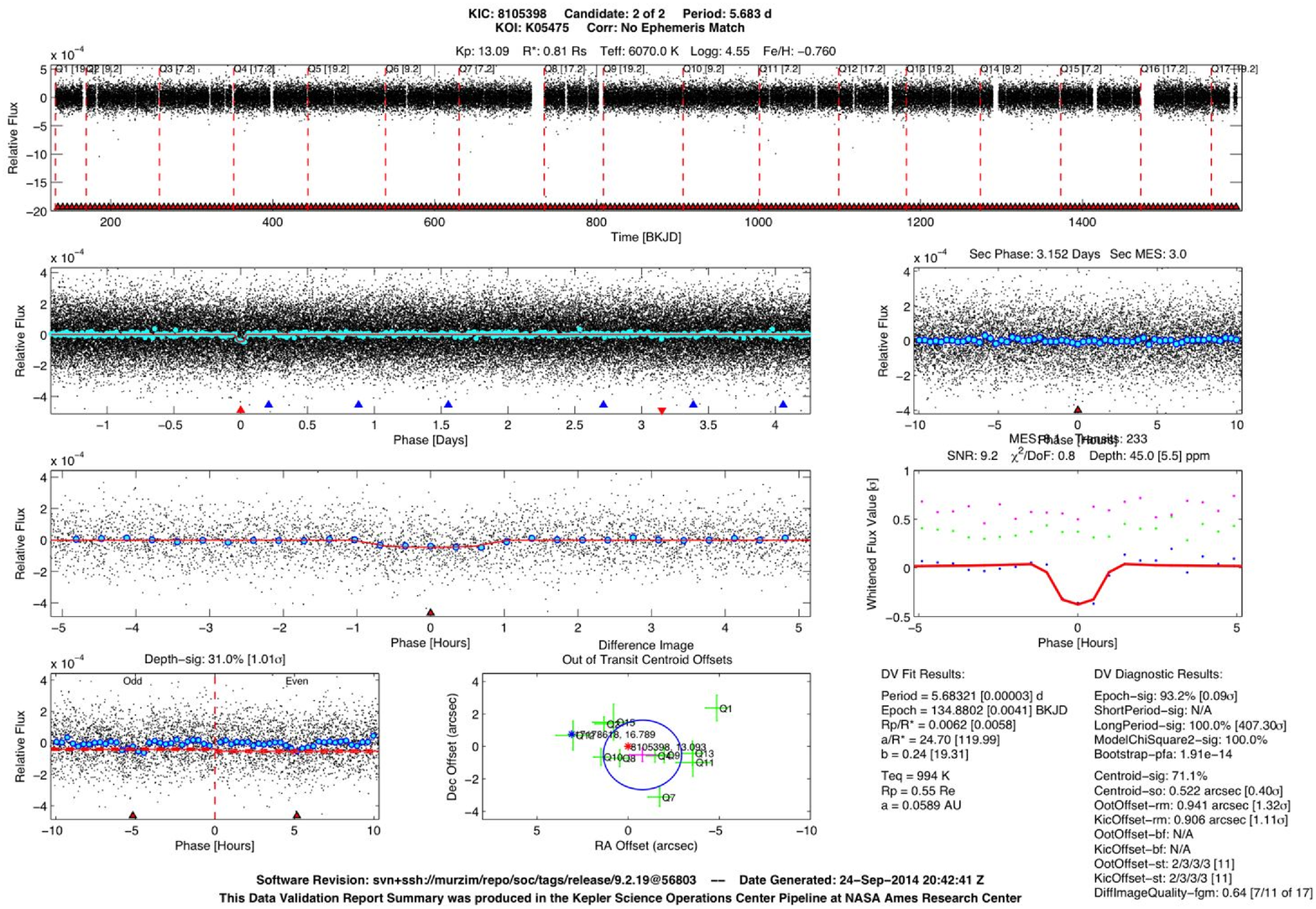}
\caption{One-page DV summary for a new candidate on KIC target 8105398.
\label{f23}}
\end{figure}

\begin{itemize}
\item Figure \ref{f16} shows the one-page summary for a new \textbf{TCE} on KIC target 8311864. This TCE has an orbital period of 384.85 days, a planet 
radius of 1.19 \rearth, and an equilibrium temperature of 221 K.  The bootstrap test measures this object as being significant at nearly the same level as
a 7.1 $\sigma$ detection in a standard normal MES distribution (an improved version of the bootstrap shows that this object is even more significant).  The true nature of many TCEs however, cannot be determined without some measure of followup observations.  
\item Figure \ref{f17} shows the one-page summary for a new \textbf{TCE} on KIC target 5094751.  This \textbf{TCE} has an orbital period of 362.5 days, a planet
radius of 1.6 \rearth, and an equilibrium temperature of 301 K.  This is KOI 123 (Kepler-109) which already has 2 confirmed planets.  
\item Figure \ref{f18} shows the one-page summary for a new \textbf{TCE} on KIC target 5531953.  This \textbf{TCE} has an orbital period of 21.91 days, a planet
radius of 0.78 \rearth, and an equilibrium temperature of 288 K.  This is KOI 1681 which already has 3 dispositioned planet candidates.
\item Figure \ref{f19} shows the one-page summary for a new \textbf{TCE} on KIC target 8120820.  This \textbf{TCE} has an orbital period of 129.22 days, a planet
radius of 1.84 \rearth, and an equilibrium temperature of 290 K. 
\item Figure \ref{f20} shows the one-page summary for a new \textbf{TCE} on KIC target 9674320.  This \textbf{TCE} has an orbital period of 317.05 days, a planet
radius of 1.66 \rearth, and an equilibrium temperature of 222 K. 
\item Figure \ref{f21} shows the one-page summary for a new \textbf{TCE} on KIC target 7100673.  This \textbf{TCE} has an orbital period of 7.24 days, a planet
radius of 0.77 \rearth, and an equilibrium temperature of 948 K. This is KOI 4032 which already has 4 dispositioned planet candidates, all with periods shorter than this one.
\item Figure \ref{f22} shows the one-page summary for a new \textbf{TCE} on KIC target 8105398.  This \textbf{TCE} has an orbital period of 224.15 days, a planet
radius of 1.71 \rearth, and an equilibrium temperature of 292 K. This is KOI 5475.01, for which a TCE was generated in the previous Q1-Q16 run \citep{pt2014} at twice the orbital period of this TCE.  This KOI was dispositioned as a false positive at the longer period due to the presence of a secondary event, but at the period of this TCE, it appears to be consistent with a transiting planet.
\item Figure \ref{f23} shows the one-page summary for a new \textbf{TCE} on KIC target 8105398.  This \textbf{TCE} has an orbital period of 5.68 days, a planet
radius of 0.55 \rearth, and an equilibrium temperature of 994 K. This is the second TCE detected on KOI 5475 and \textbf{its presence may make} the previous detection more likely to be a planet candidate rather than a false positive \citep{jack,latham}.

\end{itemize}

\section{Conclusion}
\label{s:conclusion}
The Transiting Planet Search (TPS) pipeline module was used to search photometry data for 198,675 \kepler{} targets acquired \textbf{between} May 13, 2009 \textbf{and} May 11, 2013 for science operations.  This resulted in the detection of 20,367 TCEs on 12,669 target stars.  \textbf{The distribution of TCEs presented here is qualitatively different from those obtained in a the previous search utilizing 4 years of data \citep{pt2014}: here we detect a larger number of short-period TCEs, which were detected in previous runs \citep{pt2012,pt2013}, and a much smaller number of long-period false alarms.}  The differences are believed to be due to changes made to the TPS 
algorithm, rather than to the additional flight data or changes in the data pre-processing algorithms.  The recovery rate of a set of 1,752 ``golden KOIs'' 
was $~99.1$\%.  The recovery rate across the full set of KOIs was lower however, due largely to the implementation of the bootstrap veto in TPS.  With 
a more conservative threshold and further development of this veto, some of these will be recovered.  Processing limitations prohibit us from sending too large 
a number of targets on to the final stage of the pipeline, namely, DV.  Implementing the bootstrap veto, and fixing issues with the other vetoes however,
 enables us to select the best possible \textbf{set of TCEs} for sending on.  As closeout of the \kepler{} primary mission draws near, one final pipeline development
cycle is now underway which will further improve the set of \textbf{TCEs} that comes out in the end.  The results collected, and presented here, from this 
data processing pipeline run have illuminated the key areas that will be our primary focus.

\section{Acknowledgements}
The results presented here would not be possible without the support of NASA's High End Computing Capability (HECC) Project within the 
Science Mission Directorate.  These results were generated through execution of the \textit{Kepler} Data Processing Pipeline on the NASA 
Ames supercomputer.  \textit{Kepler} was selected as the $10^{th}$ mission of NASA's Discovery Program.  Funding for this work is 
provided by NASA's Science Mission Directorate. 

\appendix
\section{The Statistical Bootstrap Test}
\label{app:bootstrap}
To search for transit signatures, TPS employs a bank of wavelet-based matched filters that form a grid on a three dimensional parameter space of transit duration, period, and phase.  The $MES$ is calculated for each template and compared to a threshold value of $\eta = 7.1 \sigma$.  Detections in TPS are made under the assumption that the underlying noise process is stationary, white, Gaussian, and uncorrelated.  When the noise deviates from these assumptions, the detection thresholds are invalid and the false alarm probability associated with such a detection may be much worse than an equivalent $MES$ for a signal embedded in white Gaussian noise.  The Statistical Bootstrap Test, or the Bootstrap, is a way of building the distribution of the null statistics from the data so that the false alarm probability can be calculated for each TCE.  

The statistic upon which detections are based is the Multiple Event Statistic ($MES$), $Z$,  whose construction is described in great detail in \cite{jmj2002}.  For simplicity here, however, the $MES$ is the output of a wavelet-based matched filter that can be written as:
\begin{equation}
Z = \sum_{i \in \mathcal{S}} \mathbb{N}(i) \slash \sqrt{ \sum_{i \in \mathcal{S}} \mathbb{D}(i) }.
\end{equation}  
Here, $\mathcal{S}$ is the set of transit times that a single period and epoch pair select out.  The $\mathbb{N}(i)$ is a correlation time series formed by correlating the whitened data to a whitened transit signal template with a transit centered at the $i'th$ time in the set $\mathcal{S}$.  The $\mathbb{D}(i)$ is the template normalization time series.  In the absence of any true signal in the data, the Probability Density Function (PDF) for the $MES$ is:
\begin{equation}
p_0(Z) = \frac{1}{\sqrt{2\pi}} \exp\left(- \frac{1}{2} Z^2 \right).
\end{equation}
The false alarm probability is then the integral of this PDF above the search threshold:
\begin{equation}
Q_0 = \int_\eta^\infty p_0(Z) \; dZ.
\end{equation}

The $MES$ distribution is a complicated function of the noise as well as the epoch, period, and transit duration.  Due to deviations from the assumed noise behavior (i.e. non-stationarity, noise artifacts, uncorrected systematics, etc.), the $MES$ distribution can be largely different across epochs, periods, or even transit durations, so it is generally not appropriate to simply sample over the full parameter space to estimate the $MES$ distribution.  The bootstrap method presented in \cite{jmj2002} sorts the data in such a way as to minimize the computation time in estimating the tail end of the distribution needed for computing the false alarm.  One weakness in such a method is that it will in general ignore any variations in the distribution across the parameter space and give a distribution which may not be representative of a detection made at a particular epoch, period, and transit duration.  The method presented here attempts to estimate the distribution for a given number of transits and transit duration.  So any non-ideal noise behavior on the same time scale of the detection will be encoded in the distribution to give a more reliable estimate of the false alarm probability.  This will also explicitly show whether the distribution deviates from its expected behavior in the regime of the detection itself, rather than in the average sense across the full parameter space.

The denominator term in the $MES$ calculation prevents a straightforward construction of the $MES$ distribution.  Since the denominator is simply a normalization term, independent of the data except through its effect on the whitening coefficients, we can begin to make progress by first ignoring it.  If the $MES$ just consisted of the numerator term, then it would be simple to form its distribution by convolution, since it is just the sum of $P$ of the same random variable.  One could simply estimate the probability density function (PDF) of the normalization time series $\mathbb{N}$ by forming its histogram, then convolve it with itself $P$ times as
\begin{equation}
\label{pdf}
PDF_{\sum_{i \in \mathcal{S}} \mathbb{N}(i)} = \Re \left\{ \mathscr{F}^{-1}\left(\left[ \mathscr{F}\left(PDF_{\mathbb{N}(n)}\right)\right]^P\right) \right\} ,
\end{equation}
where $P$ is the number of transits in the detection, $\Re$ denotes the real part of a complex number, $\mathscr{F}$ denotes the Fourier transform, and $\mathscr{F}^{-1}$ denotes the inverse Fourier transform.  This $PDF$ for the numerator of the $MES$ differs from the $PDF$ of the $MES$ only by the normalization scale factor.

The denominator term of the $MES$, the normalization sum, can be formed for each phase that the period of the detection admits.  Since some of the data are deemphasized due to various anomalies such as Earth points, spacecraft attitude tweaks, safe modes, and a host of other issues, each phase for a given period may not actually have the same number of transits, $P_j$,  as that of the detection, $P$.  For scaling purposes, however, it is important that each summation be adjusted to put it on the same scale.  So we multiply the normalization sum for each phase by a correction factor $c_j$ given by:
\begin{equation}
c_j = \left[\frac{P}{P_j}\right]^{\frac{1}{2}\sgn{\left(P-P_j\right)}} ,
\end{equation}
where $\sgn$ is the sign function.  Note however, that for some cases where a phase has a larger number of transits than the detection, but the difference is not too large, then all possible combinations of the normalization terms taken $P$ at a time are used to form a set of normalization sums for that particular phase. This improves the statistics in the end.    

When the period of the detection is large enough, the transits may not fall in every quarter of data.  When this happens, phases that use only the same quarters as were used to make the detection are admitted.  This prevents noise features from outside of the contributing quarters from skewing the distribution.  If there is little or no viable phase space for the period of the detection after applying the deemphasis weights and removing data from non-contributing quarters, then other periods are used to buoy the phase space.  These additional periods are those closest to that of the detection and are used as minimally as possible.  If there are is still no phase space out to a correlation drop of $\pm 25\%$, then the Bootstrap test is abandoned.  

Next, we divide the $PDF$ of the numerator of the $MES$ from equation (\ref{pdf}) by each of the normalization sums.  A common set of $MES$ histogram bins is then constructed by going from the min of this set up to the max.  Counts are then binned on this common axis to build the final $MES$ histogram, which is an estimate of the $MES$ PDF for the detection.  The bootstrap FAR is obtained by evaluating the complementary CDF at the TCE $MES$ value.

This version of the bootstrap depends upon separability of the correlation and normalization time series.  This is clearly not a valid assumption on many targets so we are in the process of revising the way in which the histogram is constructed.  A new method has been devised that makes no assumptions on the separability of the correlation and normalization time series.  This new method will be the main focus of an upcoming paper \citep{seaderBootstrap}.

\section{$\chi^2_{(GOF)}$ Veto}
\label{app:chisquare}
This veto measures the difference between the squared amplitude of the detector output and the squared SNR \citep{baggio,Allen}.  There
are several subtleties, described in \citet{seader2012}, associated with the construction of $\chi^2_{(2)}$.  These subtleties must 
also be taken care of in the construction of this $\chi^2_{(GOF)}$ statistic, and in what follows it is assumed that they are.
Begin with equation $(36)$ from \citet{seader2012}:
\begin{equation}
\tilde{x}_j(n) = \tilde{w}(n) + \mathcal{A} \tilde{s}_j(n),
\end{equation}
where the $\sim$ denotes a whitened vector, $n \in [1,..., N]$ with $N$ being the total number of cadences, $x(n)$ is the detector output,
$w(n)$ is the detector noise (assumed to be uncorrelated and Gaussian with zero mean and unit variance), $s_j$ is a unit amplitude
 transit signal centered at time $n=j$, and $\mathcal{A}$ is the signal amplitude.  Under the null hypothesis (i.e. no signal present)
let $\mathcal{A} \rightarrow 0$.  Choosing a particular point in the period, $T$, and epoch, $t_0$ space, selects out a set, $\mathcal{S}$, of $P$ transit centers, one for each transit, that start with the sample corresponding to the epoch $t_0$ and are spaced $T$ samples apart.  These samples form a subset of $\{n\}$, $\mathcal{S} = \{t_0,t_0+T,...,t_0+(P-1)T\}$.

Next, define the dot product of two vectors, $a(n)$ and $b(n)$ as:
\begin{equation}
\textbf{a} \cdot \textbf{b} = \sum_{n=1}^N a(n) b(n) . 
\end{equation}
Now, $\chi^2_{(GOF)}$ can be written as:
\begin{eqnarray}
\nonumber
\chi^2_{(GOF)} &=& \sum_{j=1}^P \tilde{\textbf{x}}_j \cdot \tilde{\textbf{x}}_j - \frac{{(\sum_{j=1}^P \tilde{\textbf{x}}_j \cdot \tilde{\textbf{s}}_j})^2}{\sum_{j=1}^P \tilde{\textbf{s}}_j \cdot \tilde{\textbf{s}}_j} \\
  &=& \sum_{j=1}^P \sum_{n=1}^N \tilde{x}_j(n) \tilde{x}_j(n) - \frac{({\sum_{j=1}^P \sum_{n=1}^N \tilde{x}_j(n) \tilde{s}_j(n)})^2}{\sum_{j=1}^P \sum_{n=1}^N \tilde{s}^2_j(n)}.
\end{eqnarray}
This quantitiy is $\chi^2$ distributed with $N_t - 1$ degrees of freedom, where $N_t$ is the number of in-transit cadences.  Note 
that $x_j(n)$ is zero outside of transit $j$ as discussed in \citet{seader2012}.  We threshold on this quantity in the same way 
as for the other $\chi^2$ veto as described in \citet{seader2012} and \citet{pt2013}.

\vspace{0.25cm}


\clearpage
\onecolumn
\clearpage
\clearpage
\clearpage
\clearpage
\clearpage
\clearpage
\clearpage
\clearpage
\clearpage
\clearpage
\clearpage
\clearpage
\clearpage
\clearpage
\clearpage
\clearpage
\clearpage
\clearpage
\clearpage
\clearpage
\clearpage
\clearpage

\clearpage

 commands



\end{deluxetable}

\end{document}